\documentclass[%
preprint,
tighten,
amsmath,
amssymb,
aps,
twocolumn,
]{aastex631}

\usepackage{appendix}
\usepackage{graphicx}
\usepackage{dcolumn}
\usepackage{bm}
\usepackage{color}
\usepackage{natbib}
\usepackage{comment}
\usepackage{verbatim}

\newcommand{\oversnp}{6}

\newcommand{\ds}{4}
\newcommand{\msig}{$M_{\mathrm{BH}}-\sigma_\star$}
\newcommand{\mlum}{$M_{\mathrm{BH}}-L_{\mathrm{bul}}$}
\newcommand{\mmass}{$M_{\mathrm{BH}}-M_{\mathrm{bul}}$}

\newcommand{\mbh}{$M_{\mathrm{BH}}$}
\newcommand{\inc}{$i$}
\newcommand{\ml}{$M/L_\mathrm{H}$}
\newcommand{\pa}{$\Gamma$}
\newcommand{\vsys}{$v_{\text{sys}}$}
\newcommand{\xloc}{$x_0$}
\newcommand{\yloc}{$y_0$}
\newcommand{\fw}{$f_0$}
\newcommand{\sig}{$\sigma_0$}

\newcommand{\kms}{km s$^{-1}$}

\newcommand{\mbhfullerrpgc}{\mbh\ $= (1.91\pm0.04$ [$1\sigma$ stat] $^{+0.11}_{-0.51}$ [sys])$\times 10^9$ $M_\odot$}
\newcommand{\mbhfullerrforabstractpgc}{\mbh\ $= (1.91\pm0.04$ [$1\sigma$ statistical] $^{+0.11}_{-0.51}$ [systematic])$\times 10^9$ $M_\odot$}
\newcommand{\mlfullerrforabstractpgc}{$M/L_H=1.620\pm0.004$ [$1\sigma$ statistical] $^{+0.211}_{-0.107}$ [systematic] $M_\odot/L_\odot$}
\newcommand{\mbhbothsigpgc}{\mbh\ $= (1.91 \pm 0.04$ [$1\sigma$ stat] $\pm0.11$ [$3\sigma$ stat] $^{+0.11}_{-0.51}$ [sys])$\times 10^9$ $M_\odot$}

\begin{document}

\shortauthors{Cohn et al.}

\title{ALMA gas-dynamical mass measurement of the supermassive black hole \\ in the red nugget relic galaxy PGC 11179}

\author[0000-0003-1420-6037]{Jonathan H. Cohn}
\affil{George P. and Cynthia W. Mitchell Institute for Fundamental Physics and Astronomy, Department of Physics \& Astronomy, Texas A\&M University, 4242 TAMU, College Station, TX 77843, USA}
\affiliation{Department of Physics and Astronomy, Dartmouth College, 6127 Wilder Laboratory, Hanover, NH 03755, USA}

\author[0000-0003-4612-5186]{Maeve Curliss}
\affiliation{George P. and Cynthia W. Mitchell Institute for Fundamental Physics and Astronomy, Department of Physics \& Astronomy, Texas A\&M University, 4242 TAMU, College Station, TX 77843, USA}

\author[0000-0002-1881-5908]{Jonelle L. Walsh}
\affiliation{George P. and Cynthia W. Mitchell Institute for Fundamental Physics and Astronomy, Department of Physics \& Astronomy, Texas A\&M University, 4242 TAMU, College Station, TX 77843, USA}

\author[0000-0003-2632-8875]{Kyle M. Kabasares}
\affiliation{Department of Physics and Astronomy, 4129 Frederick Reines Hall, University of California, Irvine, CA, 92697-4575, USA}

\author[0000-0001-6301-570X]{Benjamin D. Boizelle}
\affiliation{Department of Physics and Astronomy, N284 ESC, Brigham Young University, Provo, UT, 84602, USA}
\affiliation{George P. and Cynthia W. Mitchell Institute for Fundamental Physics and Astronomy, Department of Physics \& Astronomy, Texas A\&M University, 4242 TAMU, College Station, TX 77843, USA}

\author[0000-0002-3026-0562]{Aaron J. Barth}
\affiliation{Department of Physics and Astronomy, 4129 Frederick Reines Hall, University of California, Irvine, CA, 92697-4575, USA}

\author[0000-0002-8433-8185]{Karl Gebhardt}
\affiliation{Department of Astronomy, The University of Texas at Austin, 2515 Speedway, Stop C1400, Austin, TX 78712, USA}

\author[0000-0002-1146-0198]{Kayhan G\"{u}ltekin}
\affiliation{Department of Astronomy, University of Michigan, 1085 S. University Ave., Ann Arbor, MI 48109, USA}

\author[0000-0003-1693-7669]{Ak\i n Y\i ld\i r\i m}
\affiliation{Max-Planck-Institut f\"{u}r Astrophysik, Karl-Schwarzschild-Str. 1, 85748 Garching, Germany}

\author[0000-0002-3202-9487]{David A. Buote}
\affiliation{Department of Physics and Astronomy, 4129 Frederick Reines Hall, University of California, Irvine, CA, 92697-4575, USA}

\author[0000-0003-2511-2060]{Jeremy Darling}
\affiliation{Center for Astrophysics and Space Astronomy, Department of Astrophysical and Planetary Sciences, University of Colorado, 389 UCB, Boulder, CO 80309-0389, USA}

\author[0000-0002-7892-396X]{Andrew J. Baker}
\affiliation{Department of Physics and Astronomy, Rutgers, the State University of New Jersey, 136 Frelinghuysen Road Piscataway, NJ 08854-8019, USA}
\affiliation{Department of Physics and Astronomy, University of the Western Cape, Robert Sobukwe Road, Bellville 7535, South Africa}

\author[0000-0001-6947-5846]{Luis C. Ho}
\affiliation{Kavli Institute for Astronomy and Astrophysics, Peking University, Beijing 100871, China; Department of Astronomy, School of Physics, Peking University, Beijing 100871, China}

\correspondingauthor{Jonathan H. Cohn}
\email{jonathan.cohn@dartmouth.edu}

\begin{abstract}

We present 0$\farcs{22}$-resolution Atacama Large Millimeter/submillimeter Array (ALMA) observations of CO(2$-$1) emission from the circumnuclear gas disk in the red nugget relic galaxy PGC 11179.
The disk shows regular rotation, with projected velocities near the center of $400$ \kms.
We assume the CO emission originates from a dynamically cold, thin disk and fit gas-dynamical models directly to the ALMA data.
In addition, we explore systematic uncertainties by testing the impacts of various model assumptions on our results.
The supermassive black hole (BH) mass (\mbh) is measured to be \mbhfullerrforabstractpgc, and the $H$-band stellar mass-to-light ratio \mlfullerrforabstractpgc.
This \mbh\ is consistent with the BH mass$-$stellar velocity dispersion relation but over-massive compared to the BH mass$-$bulge luminosity relation by a factor of 3.7.
PGC 11179 is part of a sample of local compact early-type galaxies that are plausible relics of $z\sim2$ red nuggets, and its behavior relative to the scaling relations echoes that of three relic galaxy BHs previously measured with stellar dynamics.
These over-massive BHs could suggest BHs gain most of their mass before their host galaxies do.
However, our results could also be explained by greater intrinsic scatter at the high-mass end of the scaling relations, or by systematic differences in gas- and stellar-dynamical methods.
Additional \mbh\ measurements in the sample, including independent cross-checks between molecular gas- and stellar-dynamical methods, will advance our understanding of the co-evolution of BHs and their host galaxies.
\end{abstract}

\section{\label{intro}Introduction}

Over the past 25 years, $\sim$100 supermassive black holes (BHs) have been dynamically detected in nearby galaxies  \citep{Saglia2016}.
The masses of the BHs (\mbh) correlate with host galaxy properties like the stellar velocity dispersion ($\sigma_\star$), bulge luminosity ($L_\mathrm{bul}$), and bulge mass ($M_\mathrm{bul}$; e.g., \citealt{Kormendy1995,Ferrarese2000,Gebhardt2000,Marconi2003,Gultekin2009,Kormendy2013}).
These correlations suggest that BHs and host galaxies grow together, but the scaling relations are poorly constrained at the low- and high-mass ends.
Also, the galaxies currently populating the scaling relations are not fully representative of the diversity of assembly histories, leading to questions about how exactly BHs and host galaxies co-evolve.

A majority of \mbh\ measurements have been made by modeling stellar orbits, but recently there has been a substantial increase in the number of molecular gas-dynamical determinations with the Atacama Large Millimeter/sub-millimeter Array (ALMA; e.g., \citealt{Barth2016a,Davis2017,Boizelle2019,Nagai2019,North2019,Smith2021,Boizelle2021,Cohn2021,Kabasares2022,Nguyen2022,Ruffa2023}).
The rise in the number of molecular gas-dynamical \mbh\ measurements is due to the significant enhancements in sensitivity and angular resolution of ALMA over the previous generation of mm/sub-mm interferometers, coupled with the fact that the cold molecular gas exhibits less turbulent motion than the traditionally used ionized (e.g., \citealt{Barth2001, Walsh2010, Walsh2013}) and warm H$_2$ molecular gas (e.g., \citealt{Wilman2005,Neumayer2007,Seth2010,Scharwachter2013}) and is thus better suited for dynamical modeling.

In this paper, we analyze ALMA observations of the circumnuclear molecular gas disk in PGC 11179. PGC 11179 is a local massive, compact early-type galaxy that as yet has no \mbh\ measurement.
The object is part of a sample of 15 galaxies originally identified through the Hobby-Eberly Telescope Massive Galaxy Survey (HETMGS; \citealt{Bosch2015}) and further analyzed by \cite{Yildirim2017}.
Based on the stellar velocity dispersions \citep{Yildirim2017}, the sample of local compact early-type galaxies may have large BHs with \mbh\ up to $\sim6\times10^9\ M_\odot$ \citep{Kormendy2013,Saglia2016}.
However, the compact galaxies are distinct from the giant ellipticals and brightest cluster galaxies (BCGs) that usually host the most massive BHs in the local universe. 

\cite{Yildirim2017} find that these galaxies have small effective radii ($r_\mathrm{e} \sim 0.7-3.1$ kpc) for their stellar masses ($M_\star \sim 5.5\times10^{10} - 3.8\times10^{11}\ M_\odot$), consistent with the redshift ($z$) $\sim2$ mass$-$size relation as opposed to the $z = 0$ relation.
They display flattened shapes, fast rotation, and cuspy surface brightness profiles.
Many of the local compact galaxies also have stellar orbital distributions that show no evidence for major mergers since $z\sim2$ \citep{Yildirim2017}, uniform $\gtrsim$10 Gyr stellar ages out to several effective radii \citep{Martin2015,Mateu2017}, highly concentrated dark matter halos \citep{Buote2018,Buote2019}, and globular clusters with red color distributions \citep{Beasley2018,Kang2021}.
As such, these local compact galaxies are thought to be relic galaxies, passively evolved from $z\sim2$ massive, quiescent galaxies (``red nuggets"; e.g., \citealt{Trujillo2014,Mateu2015,Yildirim2017}) and having never grown into the present-day giant elliptical galaxies and BCGs.

Stellar-dynamical \mbh\ measurements exist for three systems in the compact galaxy sample: NGC 1277, NGC 1271, and Mrk 1216 \citep{Bosch2012,Emsellem2013,Yildirim2015,Graham2016a,Walsh2015,Walsh2016,Walsh2017,Krajnovic2018}.
The \mbh\ in NGC 1277 has also been studied with molecular gas dynamics \citep{Scharwachter2016}.
Although the bulge fractions for the compact galaxies have been debated (e.g., \citealt{SavorgnanGraham2016,Graham2016b}), the objects remain positive outliers from the \mlum\ and \mmass\ scaling relations, even when global properties are treated as ``bulge" luminosities and masses, while falling on the \msig\ relation (\citealt{Walsh2015,Walsh2016,Walsh2017}).
If the local compact galaxies with over-massive BHs are relics of red nuggets, this could be evidence that BHs complete most of their growth by $z\sim2$, after which typical massive elliptical galaxies accrete more stellar mass with minor/intermediate dry mergers, growing their outskirts without significantly increasing their central $\sigma_\star$ or significantly feeding their BHs (e.g., \citealt{Dokkum2010,Hilz2013}).

Complicating the picture, the first ALMA molecular gas-dynamical \mbh\ for one of the local compact galaxies, UGC 2698, was found to be consistent with all three of the scaling relations \citep{Cohn2021}.
However, UGC 2698 may have undergone an intermediate-to-major merger after $z\sim2$ \citep{Yildirim2017} and could thus represent a more intermediate evolutionary step compared to the other galaxies in the sample, making it a less pristine relic.
Alternatively, the UGC 2698 result could indicate that there is more scatter in the scaling relations than previously thought, either intrinsically or due to differences between stellar-dynamical and molecular gas-dynamical measurement methods \citep{Cohn2021}.
Therefore, obtaining more BH mass measurements --- with both stellar and molecular gas-dynamical methods --- for the local compact galaxy sample is required to discriminate between the two explanations.
Such measurements will help populate the poorly sampled high-mass end of the scaling relations, while simultaneously investigating likely relic galaxies with unique growth histories.
In addition to PGC 11179, which is studied here, there remain six local compact galaxies with dust disks suitable for ALMA-based dynamical \mbh\ measurements.

The structure of the paper is as follows.
We present the Hubble Space Telescope (HST) and ALMA observations for PGC 11179 in \S\ref{observations} and discuss our dynamical model and parameter optimization in \S\ref{model}.
We present the results of our model, including the inferred BH mass, in \S\ref{results}.
In \S\ref{discussion}, we estimate the BH gravitational sphere of influence (SOI), compare our results to other dynamical \mbh\ measurements in the local compact galaxy sample, and discuss the significance of our results with respect to understanding BH$-$galaxy co-evolution.
In \S\ref{conclusions}, we summarize our conclusions.
Throughout this work, we use an angular diameter distance to PGC 11179 of 89 Mpc, where 431 pc equals 1\arcsec.
This estimate is based on a $\Lambda$CDM cosmology with H$_0 = 73$ km s$^{-1}$ Mpc$^{-1}$, $\Omega_M = 0.31$, and $\Omega_{\Lambda} = 0.69$, adopting the Hubble Flow distance from the Virgo + Great Attractor + Shapley Supercluster infall model \citep{Mould2000} from the NASA/IPAC Extragalactic Database\footnote[10]{\url{https://ned.ipac.caltech.edu/}}.
The BH mass we measure scales linearly with the assumed distance to the galaxy.

\section{\label{observations}Observations}

Molecular gas-dynamical \mbh\ measurements require the characterization of the host galaxy's stellar light profile, as stars contribute to the gravitational potential, and high-resolution observations of emission in the circumnuclear gas disk.
Here, we discuss our HST Wide Field Camera 3 (WFC3) imaging and ALMA observations.

\subsection{HST Imaging\label{hst}}

PGC 11179 was observed with HST WFC3 on 2013 August 7 as part of program GO-13050 (PI: van den Bosch) in the IR/F160W ($H$-band) and UVIS/F814W ($I$-band) filters.
The $H$-band observations consisted of three dithered full-array exposures along with four dithered short sub-array exposures.
The sub-array sequence better samples the point spread function (PSF) and avoids saturating the nucleus.
For the $I$-band imaging, three dithered full-array exposures were taken.

\begin{figure*}
\includegraphics[width=\textwidth]{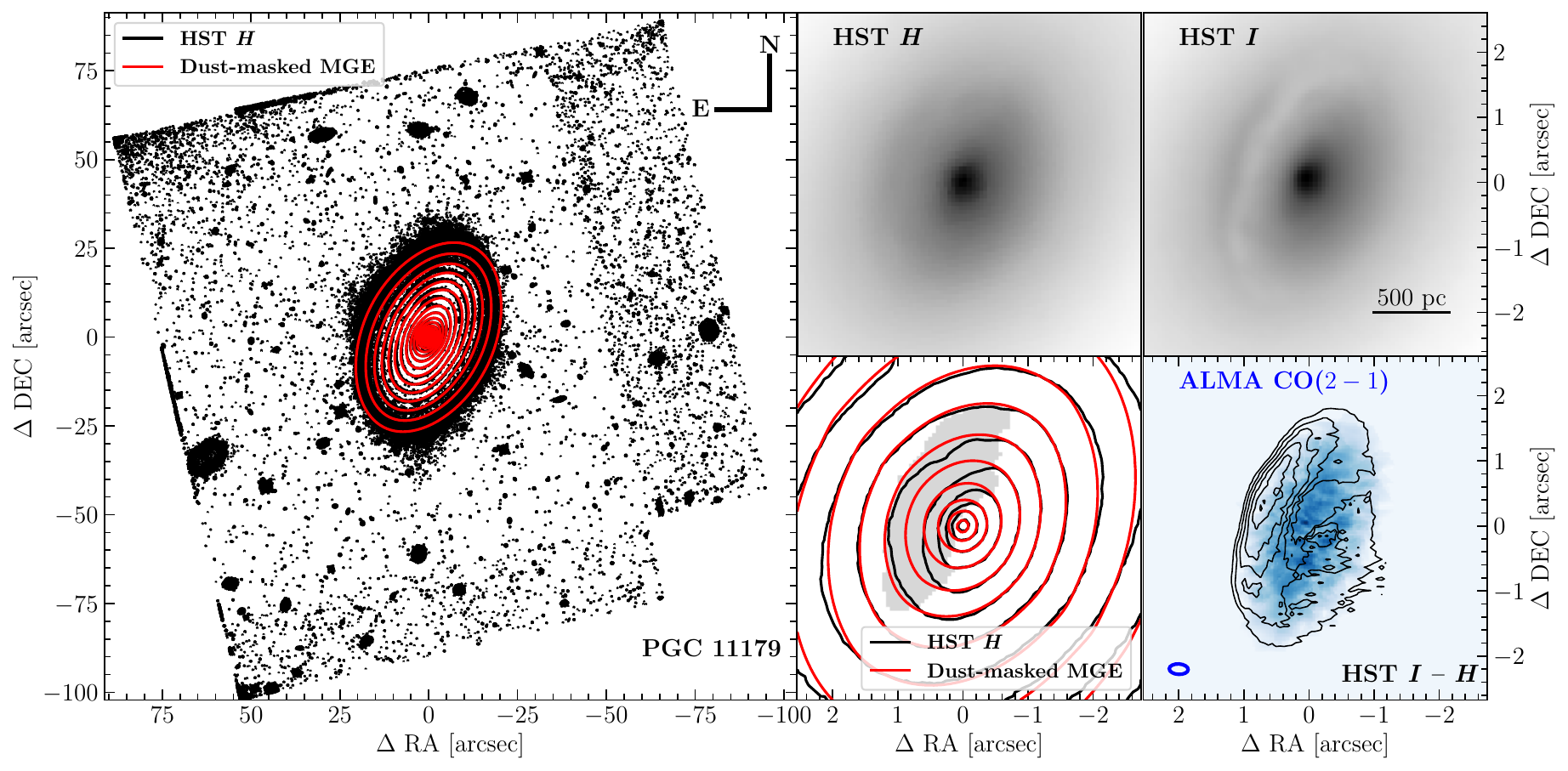}
\caption{
(Left) HST F160W ($H$-band) image (black contours) with the dust-masked MGE model (red contours) overlaid.
(Middle, top) The central $5\arcsec$ of the HST $H$-band image.
(Right, top) The inner $5\arcsec$ of the HST F814W ($I$-band) image.
The dust disk (with a diameter of $\sim$3$\farcs{9}$) is clearly visible in the $I$-band image, particularly to the east/northeast of the nucleus.
(Middle, bottom) Contours of the inner $5\arcsec$ of the HST $H$-band image (black) and the dust-masked MGE (red).
The gray shaded region was masked for the dust-masked MGE fit.
The black contours are asymmetric on the east/northeast side of the disk, indicating that dust clearly diminishes the stellar surface brightness within the masked region.
The asymmetry is no longer present in the red model contours.
(Right, bottom) The inner $5\arcsec$ of the ALMA CO(2$-$1) emission (blue) with the HST $I-H$ image (black contours) overplotted.
The blue ellipse in the lower left of the panel displays the size of the synthesized ALMA beam.
The dust disk and CO($2-1$) emission are co-spatial, with the dust disk's diameter ($\sim$3$\farcs{9}$) approximately the same size as the CO disk.
}
\label{fig_hst_pgc}
\end{figure*}

We use the processed HST $H$-band data from \citet{Yildirim2017}, which was run through the {\tt calwf3} pipeline and {\tt AstroDrizzle} \citep{Gonzaga2012} to produce combined, cleaned, distortion-corrected images.
The final $H$-band image has a pixel scale of $0\farcs06$ pixel$^{-1}$, with a field of view (FOV) of $2\farcm7\times2\farcm6$ and an exposure time of 1354.5 s.
We also drizzle the $I$-band image to the same pixel scale as the $H$-band image in order to construct an $I-H$ map.
The final $I$-band image has an exposure time of 495.0 s.
Prior to constructing the color map, we degrade the sharper $I$-band image to match the resolution of the $H$-band image.
The $H$, $I$, and $I-H$ images are displayed in Figure \ref{fig_hst_pgc}.
  
The HST $I$-band image clearly shows a circumnuclear dust disk, spanning a diameter of $\sim$3$\farcs{9}$, and the $H$-band image displays an asymmetry in the surface brightness along the east/northeast side of the nucleus, also indicating the presence of dust attenuation.
We find a maximum color excess $\Delta(I-H)\sim$0.6 mag approximately $1\farcs{1}$ to the east/northeast of the nucleus.
This color excess is measured relative to the median $I-H=1.8$ mag just beyond the disk region.
Throughout this paper, we use Vega magnitudes.

\subsubsection{Modeling the Galaxy Surface Brightness\label{mge}}

We parameterize the surface brightness profile of PGC 11179 using Multi-Gaussian Expansions (MGEs), which accurately reproduce the light profiles of early-type galaxies \citep{Emsellem1994,Cappellari2002}.
The sum of two-dimensional (2D) Gaussians is fit to the $H$-band image using {\tt GALFIT} \citep{Peng2010}, with initial parameter guesses set to results from a prior run of the {\tt mgefit} package with regularization \citep{Cappellari2002}.
We constrain Gaussian components in {\tt GALFIT} to have identical position angles (PAs) and centroid locations, and during the fit, the PSF is taken into account.
The PSF comes from Tiny Tim \citep{Krist2004}, drizzled and dithered in a manner identical to the galaxy observations.
Following \citet{Cohn2021}, our goal is to produce multiple MGEs for the galaxy as a way to assess the impact of dust on the inferred \mbh. 

As a starting point, the $H$-band image is masked to exclude foreground stars, galaxies in the image, and detector artifacts.
This resulting MGE is referred to as the ``original MGE" for the remainder of the paper. We find the original MGE has nine components, each with a PA of $154.222^\circ$ east of north.
The $\sigma^\prime$ (projected Gaussian width) values range from $0\farcs{088}$ to $23\farcs{505}$ and $q^\prime$ (projected axis ratio) is between 0.526 and 1.000.
This MGE, however, yields large residuals in the innermost $\sim$5$\arcsec$, rising to $\sim$20\%.
Given the poor residuals and the substantial dust attenuation visible in the $H$-band image (see the highly asymmetric image contours in Figure \ref{fig_hst_pgc}), it is not feasible to use this original MGE in the \mbh\ systematic uncertainty analysis, so below introduce alternative MGEs.

Next, we generate two MGEs based on an $H$-band image with the most dust-contaminated regions masked.
We construct an initial dust mask, following \citet{Cohn2021}, masking pixels redder than $I-H = 2.25$.
An initial MGE is fit with {\tt GALFIT} to the $H$-band image using the mask based on this color cut.
We then inspect the residuals near the nucleus, extend the mask to exclude pixels with large residuals, and re-fit the MGE.
After the first iteration of this process, the MGE is a reasonable fit to the image, with $\lesssim$10\% residuals at the center.
We refer to this resulting MGE as the ``minimally masked MGE" for the remainder of the paper.
We further repeat the process, extending the mask based on the residuals and re-fitting the MGE, and iterate until the residuals are $\lesssim$5\%.
For the rest of the paper, this resulting MGE is called the ``dust-masked MGE," and any time we refer to the dust mask, we mean the final mask generated through the full iterative approach. 

\begin{figure}
\includegraphics[width=0.47\textwidth]{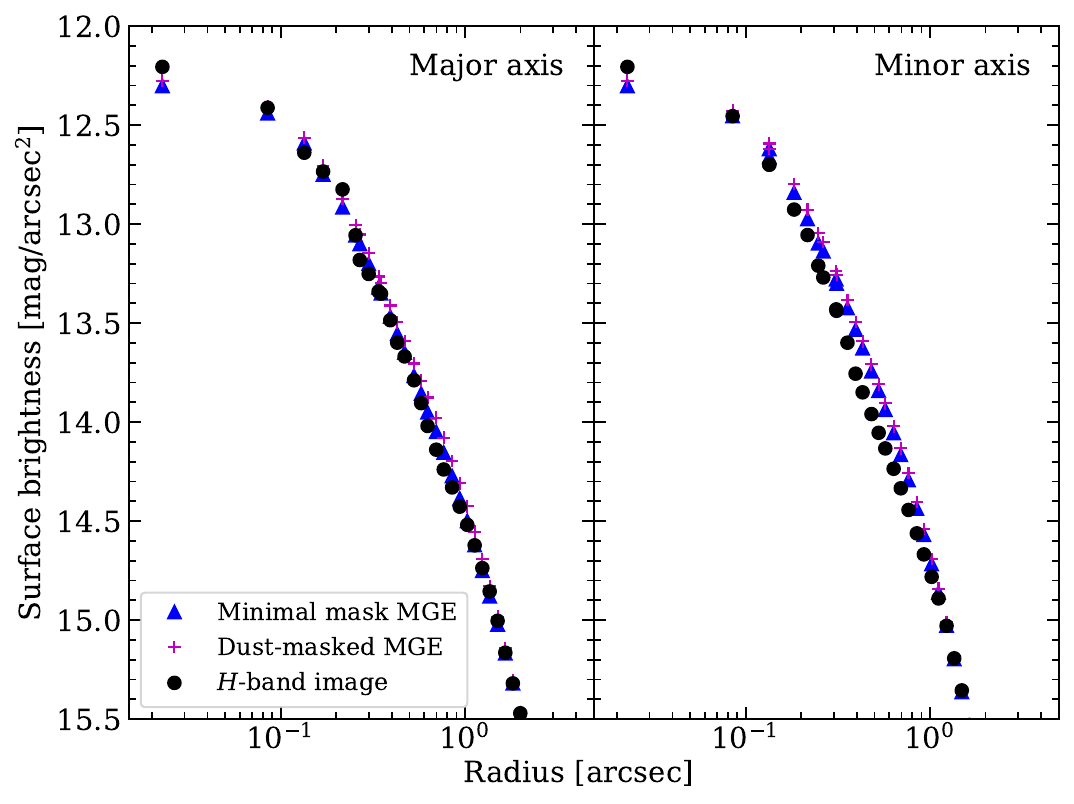}
\caption{The central surface brightness profiles of PGC 11179 along the major and minor axes as a function of projected radius based on the $H$-band image (black circles), the minimally masked MGE (blue triangles), and the dust-masked MGE (pink pluses).
We use the dust-masked MGE in our fiducial dynamical model.
}
\label{sbprof_pgc}
\end{figure}

Thus, the minimally masked MGE is based on a dust mask that is less aggressive than the final mask used to produce the dust-masked MGE.
The minimally masked MGE consists of eleven Gaussians with a PA of 154.079$^\circ$ east of north.
The components have $0\farcs{074} \leq \sigma^\prime \leq 20\farcs{801}$ and $q^\prime$ ranging from 0.539 to 0.998.
The parameters of the dust-masked MGE are given in Table \ref{tab_mgepgc} and the contours of the MGE and image are plotted in Figure \ref{fig_hst_pgc}.
We display the major- and minor-axis surface brightness profiles in the central region of the galaxy extracted from the data, the minimally masked MGE, and the dust-masked MGE in Figure \ref{sbprof_pgc} in order to highlight the differences between the models.

Finally, we construct another model of the surface brightness by estimating a dust correction, adjusting the $H$-band image, and fitting an MGE with {\tt GALFIT}.
Following \citet{Viaene2017} and \citet{Boizelle2019}, we assume the dust is in a thin, inclined disk within the galaxy.
The galaxy is taken to be oblate axisymmetric and we adopt the same inclination angle for the gas disk, dust disk, and stellar component of the galaxy, corresponding to 60.0$^\circ$ based on initial gas-dynamical models (see \S\ref{results}).
The original MGE is deprojected given the inclination angle, and we determine the fraction of starlight in front of and behind the inclined dust disk as a function of spatial location.
In our model, the stellar light in front of the disk is unaffected by dust while the light originating from behind the disk is obscured by simple screen extinction.
Using Equations 1 and 2 in \citet{Boizelle2019}, we generate model color excess curves as a function of intrinsic dust extinction, $A_V$, with a standard Galactic $R_V=3.1$ extinction curve \citep{Rieke1985} to convert between $A_V$ and $A_H$ or $A_I$.

As described in \citet{Boizelle2019}, the model color excess curve increases with intrinsic $A_V$ up to a turnover point, after which the color excess decreases.
Thus, when comparing the observed $\Delta(I-H)$ to the model color excess curve, there are generally two possible values of intrinsic $A_V$.
Following previous precedent (e.g., \citealt{Boizelle2019,Boizelle2021,Cohn2021,Kabasares2022}), we adopt the lower intrinsic $A_V$ value and convert to $A_H$.
While the comparison between observed $\Delta(I-H)$ and the model color excess curve is done for every spatial location, we do not attempt to apply a pixel-by-pixel correction to the image.
Properly determining the intrinsic stellar surface brightness may require radiative transfer models that incorporate the disk thickness, geometry, extinction and dust scattering within the disk.
Instead, we calculate the median $A_H$ within the dust mask, finding $A_H=0.3$, and use this single extinction value to perform our dust correction.

To create a dust-corrected image, we follow a process similar to those of \citet{Boizelle2019} and \citet{Kabasares2022}.
We first fit a 2D Nuker model \citep{Faber1997} with {\tt GALFIT} to the central $5\arcsec\times5\arcsec$ region of the dust-masked $H$-band image, including PSF blurring.
The Nuker profile is a double power-law characterized by inner slope $\gamma$, outer slope $\beta$, and sharpness of transition $\alpha$ at a break radius $r_b$.
From the fit, we find $\alpha=2.93$, $\beta=1.89$, $\gamma=0.85$, and $r_b=1.11\arcsec$ ($\sim$477 pc).
We then correct the pixel values of the $H$-band image within the dust mask using the previously determined $A_H$ value of $0.3$.
Holding all other Nuker parameters fixed to the prior best-fit values, we refit the adjusted $H$-band image allowing $\alpha$ and $\gamma$ to vary and determine $\gamma=0.84$ and $\alpha=0.43$.
We replace the dust-masked region of the original $H$-band image with the pixel values of the Nuker model.
With this method, the dust-corrected image has a smoothly varying light distribution.
Lastly, we fit an MGE to the dust-corrected $H$-band image using {\tt GALFIT}.
Hereafter, this MGE will be referred to as the ``dust-corrected MGE".
The dust-corrected MGE is composed of eleven Gaussians, with $0\farcs{065} \leq \sigma^\prime \leq 20\farcs{362}$, $0.524 \leq q^\prime \leq 1.000$, and a PA of 154.137$^\circ$ east of north.
The dust-corrected MGE has residuals at the $\lesssim$5\% level relative to the data.
Although the major- and minor-axis surface brightness profiles from the dust-corrected MGE are not shown in Figure \ref{sbprof_pgc}, they are very similar to those for the dust-masked MGE.

Ultimately, we use the minimally masked MGE, the dust-masked MGE, and the dust-corrected MGE when constructing dynamical models to assess the effect of dust on the inferred \mbh\ in \S \ref{error_pgc}.
Following \citet{Cohn2021}, we use the dust-masked MGE as the fiducial model.

\begin{deluxetable}{cccc}[ht]
\tabletypesize{\small}
\tablecaption{Dust-masked MGE parameters}
\tablewidth{0pt}
\tablehead{
\colhead{$j$} & 
\colhead{$\log_{10}(I_{H,j})$ [$L_{\odot}$ pc$^{-2}$]} & 
\colhead{$\sigma_j^\prime$ [arcsec]} & 
\colhead{$q_j^\prime$}
\\[-1.5ex]
\colhead{(1)} & 
\colhead{(2)} & 
\colhead{(3)} & 
\colhead{(4)}
}
\startdata
1 & 5.361 & 0.072 & 0.852 \\
2 & 4.683 & 0.216 & 0.922 \\
3 & 4.315 & 0.522 & 0.875 \\
4 & 3.993 & 0.938 & 0.673 \\
5 & 3.730 & 1.265 & 0.830 \\
6 & 3.481 & 2.432 & 0.706 \\
7 & 2.934 & 4.201 & 0.544 \\
8 & 2.706 & 7.756 & 0.547 \\
9 & 1.924 & 12.637 & 0.629 \\
10 & 1.156 & 17.340 & 0.532 \\
11 & 1.086 & 20.732 & 1.000
\enddata
\begin{singlespace}
  \tablecomments{Fiducial MGE parameters found by fitting the dust-masked HST $H$-band image of PGC 11179.
  Column (1) lists the MGE component.
  Column (2) displays the component's central surface brightness, calculated using an absolute $H$-band magnitude for the Sun of 3.37 mag \citep{Willmer2018} and 0.096 mag for Galactic extinction toward PGC 11179 \citep{Schlafly2011}.
  Column (3) shows the Gaussian component's projected dispersion along the major axis, and Column (4) lists the component's axis ratio.
  Primed parameters are projected values.
  Each component has a PA of $154.104^\circ$ east of north.}
\end{singlespace}
\label{tab_mgepgc}
\end{deluxetable}

\subsection{ALMA Data\label{alma}}

We obtained Cycle 4 ALMA band 6 observations as part of Program 2016.1.01010.S (PI: Walsh) in the C40$-$6 configuration with minimum and maximum baselines of 16.7 m and 2600 m.
PGC 11179 was observed on 2017 July 8 using a single pointing with one spectral window centered at 225.258 GHz, corresponding to the redshifted frequency of the 230.538 GHz $^{12}$CO(2$-$1) line, and two continuum spectral windows with average frequencies of 227.402 GHz and 240.766 GHz.
The on-source exposure time was 25.3 minutes.
Dust emission is detected in the continuum spectral windows, but we focus on the CO spectral window in this work.

We use Common Astronomy Software Applications (CASA) version 4.7.2 to process the ALMA data.
We employ a \texttt{TCLEAN} deconvolution with Briggs weighting ($r = 0.5$; \citealt{Briggs1995}) and perform $uv$-plane continuum subtraction using the emission-free channels.
The resultant synthesized beam has a full width at half maximum (FWHM) of $0\farcs{29}$ along the major axis and $0\farcs{16}$ along the minor axis, with a geometric mean of $0\farcs{22}$ (92.7 pc) and a PA of $86.97^{\circ}$ east of north.
The observations were flux-calibrated with the ALMA standard quasar J$0006-0623$, and we assume a $10\%$ uncertainty in the absolute flux calibration at this frequency \citep{Fomalont2014}.

The final PGC 11179 data cube has a pixel scale of $0\farcs{03}$ pixel$^{-1}$ and 65 frequency channels that are 14.76 MHz wide, which corresponds to $\sim$19.63 \kms\ at the redshifted CO($2-1$) frequency.
We detect CO emission in channels 10 through 53, which correspond to recessional velocities $cz= 6409.2 - 7253.4$ \kms.
The emission-free regions of the data cube have a root-mean-square (rms) noise level of 0.4 mJy beam$^{-1}$ channel$^{-1}$.

\subsubsection{Emission-Line Properties\label{emission}}

\begin{figure*}
\includegraphics[width=\textwidth]{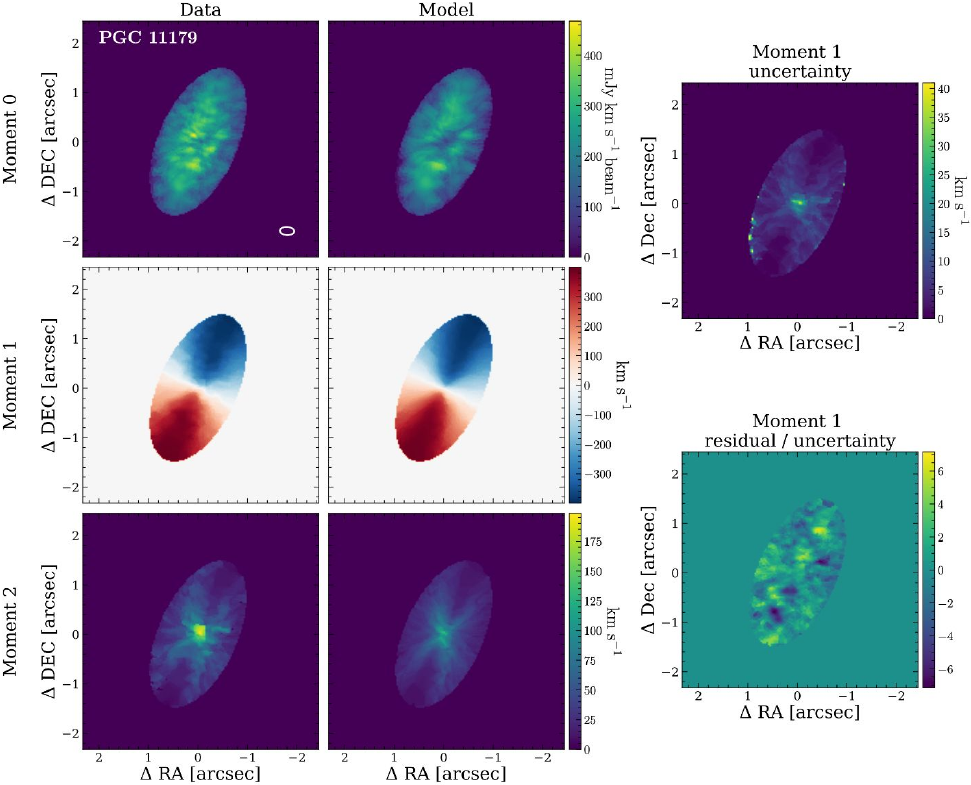}
\caption{Zeroth, first, and second moments from the PGC 11179 ALMA data cube (left) and the best-fit model (middle) within the fiducial elliptical fitting region.
The synthesized ALMA beam is shown as a white ellipse in the upper left panel.
The uncertainty of the moment one map, calculated from a 1000-iteration Monte Carlo simulation, is shown in the upper right, and the residual (data$-$model) of the first moment normalized by the uncertainty is shown in the bottom right.
These maps are constructed on the original $0\farcs{03}$ pixel$^{-1}$ ALMA pixel scale.
All panels are linearly mapped to colors given by the scale bars, with data and model maps for each moment shown on the same scale.
Note that our dynamical models are fit directly to the data cube, not to these moment maps.
The maps here are thus not used in the fit, and are simply extracted from the data cube and the best-fit model cube.
}
\label{fig_fiducial_moments_pgc}
\end{figure*}

Figure \ref{fig_fiducial_moments_pgc} displays spatially resolved maps of the zeroth, first, and second moments of the ALMA observations of PGC 11179.
These moments correspond to integrated CO(2$-$1) emission, projected line-of-sight velocity ($v_{\mathrm{los}}$), and projected line-of-sight velocity dispersion ($\sigma_{\mathrm{los}}$), respectively.
Pixels that contain no discernible CO emission are masked when building these maps, and the maps are further masked by the elliptical fitting region used during the dynamical modeling (see \S \ref{model}).
We also show the uncertainty in the observed $v_{\mathrm{los}}$, which we determine using a Monte Carlo simulation.
We generate 1000 mock data cubes with pixel values drawn randomly from a normal distribution centered on the observed pixel value, with the width of the distribution set by the standard deviation of an emission-free region in each channel of the data cube.
After each iteration we construct the first moment map, and we take the standard deviation of the resultant 1000 maps as the uncertainty.

The CO emission traces a regularly rotating disk, in which the southeast side is redshifted and the northwest side is blueshifted, with line-of-sight velocities peaking at $\pm400$ \kms.
The second moment map reaches $197$ \kms\ and displays an ``X" shape due to rotational broadening and beam smearing in areas with steep velocity gradients.
The molecular gas disk is co-spatial with the dust disk, as seen in Figure \ref{fig_hst_pgc}.

\begin{figure*}
\includegraphics[width=\textwidth]{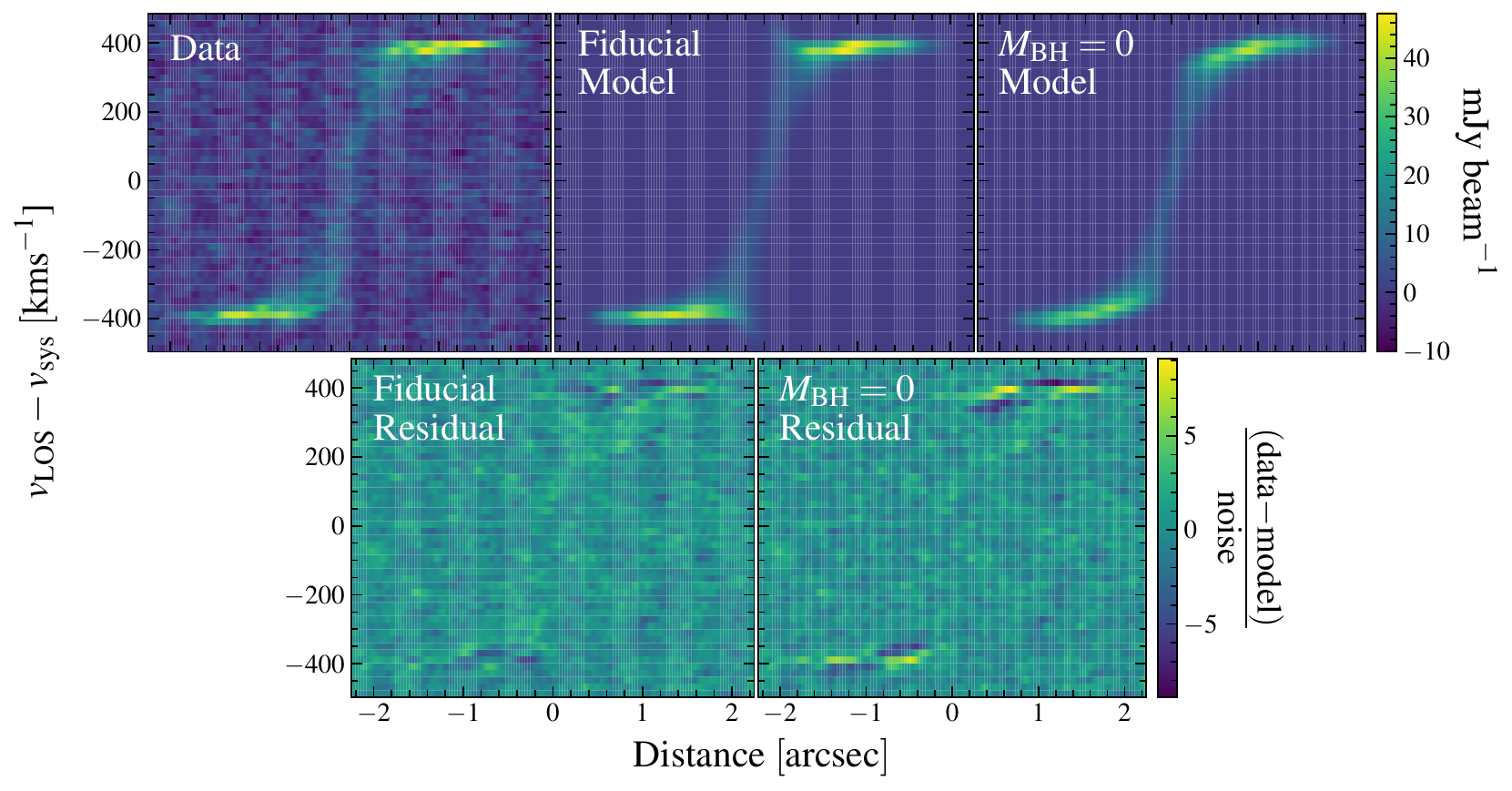} 
\caption{(Top row) PVDs built from the observed data cube (left), the best-fit fiducial model (middle), and the best-fit model with \mbh\ fixed to 0 (right).
(Bottom row) PVD residuals for the best-fit fiducial model (left) and the best-fit model with \mbh\ fixed to 0 (right).
The residuals are normalized by the noise, calculated as the standard deviation of an emission-free region of the observed data cube's PVD.
The PVDs were extracted along the major axis of the disk at a PA of 335.7$^\circ$ east of north, with a width of $0\farcs{22}$ that matches the geometric mean of the synthesized beam dimensions.
The panels are mapped linearly to the colors of the scale bar.
Our dynamical models are fit directly to the data cube, not to these PVDs.
The PVDs are not used in the fit, and are simply extracted from the data cube and the best-fit model cube.
}
\label{fig_pvds}
\end{figure*}

We construct position-velocity diagrams (PVDs), extracting the flux along the major axis at an angle of $335.7^\circ$ measured east of north to the blueshifted side of the disk.
This angle corresponds to the best-fit gas disk PA from our dynamical model.
We set the extraction width to the geometric mean of the ALMA beam dimensions.
The observed PVD is presented in Figure \ref{fig_pvds}, showing continuous emission across the full range of velocities in the gas disk.

The total CO(2$-$1) flux is $15.37 \pm0.08$ (stat) $\pm 1.54$ (sys) Jy km s$^{-1}$.
We estimate the CO(2$-$1) luminosity ($L^\prime_{\mathrm{CO(2-1)}}$) from the observed flux \citep{Carilli2013} and convert to a CO(1$-$0) luminosity ($L^\prime_{\mathrm{CO(1-0)}}$) by taking $R_{21} \equiv L^\prime_{\text{CO}(2-1)}/L^\prime_{\text{CO}(1-0)} = 0.7$ (for an excitation temperature of $\sim5-10$ K; e.g., \citealt{Lavezzi1999}).
To convert to an H$_2$ mass, we assume the CO-to-H$_2$ conversion factor is $\alpha_{\text{CO}} = 3.1\ M_\odot$ pc$^{-2}$ (K km s$^{-1}$)$^{-1}$ \citep{Sandstrom2013}.
The H$_2$ gas mass is $M_{\text{H}_2} = \alpha_{\text{CO}} L^\prime_{\mathrm{CO(1-0)}}$, from which we derive a total gas mass, adopting a helium mass fraction $f_{\text{HE}} = 0.36$.
Thus, the total gas mass is $M_{\text{gas}} = M_{\text{H}_2} (1+f_{\text{HE}})$.
We find a total gas mass of $(4.39\pm0.02\ \mathrm{[stat]} \pm0.44\ \mathrm{[sys]})\times10^8\ M_\odot$.
We note that there may be larger systematic uncertainties in the total gas mass due to the assumed $\alpha_{\text{CO}}$, as the $\alpha_{\text{CO}}$ we reference was calibrated for spiral galaxy disks and thus may not apply directly to the disks in early-type galaxies.
Nevertheless, this gas mass is approximately the same order of magnitude as those found in UGC 2698 \citep{Cohn2021} and other early-type galaxies with CO detections \citep{Boizelle2017,Ruffa2019}.

\section{\label{model}Dynamical Modeling}

Following \citet{Barth2016b}, \citet{Boizelle2019}, and \citet{Cohn2021}, we assume the molecular gas experiences circular rotation in a thin disk.
At each radius in the disk, we calculate the circular velocity ($v_c$) relative to the systematic velocity ($v_\mathrm{sys}$) based on the enclosed mass due to the BH and the stars.
We treat the gas mass as negligible, as well as the dark matter mass, which is thought to be insignificant on the scale of the gas disk.
The enclosed stellar mass is determined by deprojecting the dust-masked MGE, assuming an oblate axisymmetric shape and an inclination angle that matches that of the gas disk (\inc), and multiplying by the stellar mass-to-light ratio (\ml).
The circular velocities are generated on a grid that is oversampled relative to the ALMA data cube by a factor of $s = $ \oversnp\ in both $y$ and $x$ and converted to $v_\mathrm{los}$ using \inc\ and the position angle of the gas disk (\pa).

We adopt a Gaussian function as the intrinsic line profile shape, centered on $v_\mathrm{los}$ at each point on the model grid and given an intrinsic width ($\sigma_\mathrm{turb}$).
We take $\sigma_\mathrm{turb}$ to be a constant (\sig) throughout the disk.
The line profiles are weighted by an estimate of the intrinsic CO flux distribution, formed by deconvolving the zeroth moment map with with the ALMA beam.
We use 10 iterations of the \texttt{scikit-image} \citep{Walt2014} package's \texttt{lucy} task \citep{Richardson1972,Lucy1974} for the deconvolution.
The flux in each pixel is divided evenly among the $s \times s$ subpixels, and we also use a scale factor of order unity, \fw, to account for slight mismatches in normalization between the model and data line profiles.
We build the Gaussian line profiles on the observed ALMA data cube's frequency axis, with a channel spacing of 14.76 MHz.

By summing the model line profiles within each set of $s \times s$ subpixels, we produce an intrinsic model cube with the same pixel scale as the ALMA cube.
Next, each frequency slice of the model cube is convolved with the synthesized ALMA beam.
Finally, we compare the model and data cubes directly, after down-sampling both in bins of $\ds\times\ds$ spatial pixels to mitigate correlated noise \citep{Barth2016b}.
The comparison between the model and data cubes is done within a fixed elliptical region in each velocity channel, constructed to encompass almost all of the CO emission while excluding excess noise.
We use a region with a semi-major axis of $r_{\text{fit}} = 1\farcs{6}$, an axis ratio $q_{\text{ell}} = 0.50$, and a position angle \pa$_{\text{ell}} = 335.7^\circ$ east of north.
The fitting region spans 50 velocity channels, corresponding to $|v_\mathrm{los}-v_\mathrm{sys}|\lesssim 490$ km s$^{-1}$, and includes 14050 data points, with 14041 degrees of freedom.

We optimize the nine free parameters in our fiducial model [\mbh, \ml, \inc, \pa, \vsys, \sig, the BH location (\xloc, \yloc), and \fw] with the nested sampling code \textbf{{\tt dynesty} \citep{Speagle2020}}, adopting a likelihood of $L \propto \exp(-\chi^2/2)$, where $\chi^2 = \sum_{i} ((d_i - m_i)^2/\sigma_i^2)$.
The down-sampled data and model points in each channel are $d_i$ and $m_i$, respectively, and $\sigma_i$ is the noise per velocity channel, estimated as the standard deviation in an emission-free region of the down-sampled ALMA data cube.
We use the Dynamic Nested Sampler in {\tt dynesty}, with a threshold of 0.02 and 250 live points.
The results are robust to our choice of live points and threshold.
We adopt flat priors (listed in Table \ref{tab_fiducial_pgc}) and sample the parameters uniformly in linear space, except for \mbh, which we sample in logarithmic space.
We report the $68\%$ ($99.7\%$) confidence intervals of the parameter posterior distributions as the $1\sigma$ ($3\sigma$) statistical uncertainties.
Additional details about the modeling can be found in \citet{Cohn2021}.

\section{\label{results}Modeling Results}
Below, we discuss the results from fitting our rotating thin-disk models to the ALMA data cube of PGC 11179.
We then discuss tests to explore the systematic uncertainties in the resultant BH mass.

\subsection{Fiducial Model Results \label{p11179results}}
For our fiducial model of PGC 11179, we find \mbh\ = $(1.91\pm0.04\ [\pm0.12]) \times 10^9\ M_\odot$, \ml\ $ = 1.620\pm0.004\ [\pm0.012]\ M_\odot/L_\odot$ (with $1\sigma$ and [$3\sigma$] uncertainties), $\chi^2 = 20476.4$, and reduced $\chi^2_\nu = 1.458$.
The remaining best-fit parameters are listed in Table \ref{tab_fiducial_pgc}.
For comparison, simple stellar population (SSP) models \citep{Vazdekis2010} predict \ml\ $\sim$ $1.2$ $M_\odot/L_\odot$ for a \cite{Kroupa2001} Initial Mass Function (IMF) with a $\sim12-13.5$ Gyr stellar age and metallicity $\sim$0.1 dex above solar.
The age and metallicity are selected here based on the measurements for PGC 11179 in \cite{Yildirim2017}.
This is consistent with the suggestion of a more bottom-heavy IMF existing in the central regions of massive ETGs with large velocity dispersions (e.g., \citealt{Martin2015a,Martin2015,LaBarbera2019}).

\begin{figure*}
\includegraphics[width=\textwidth]{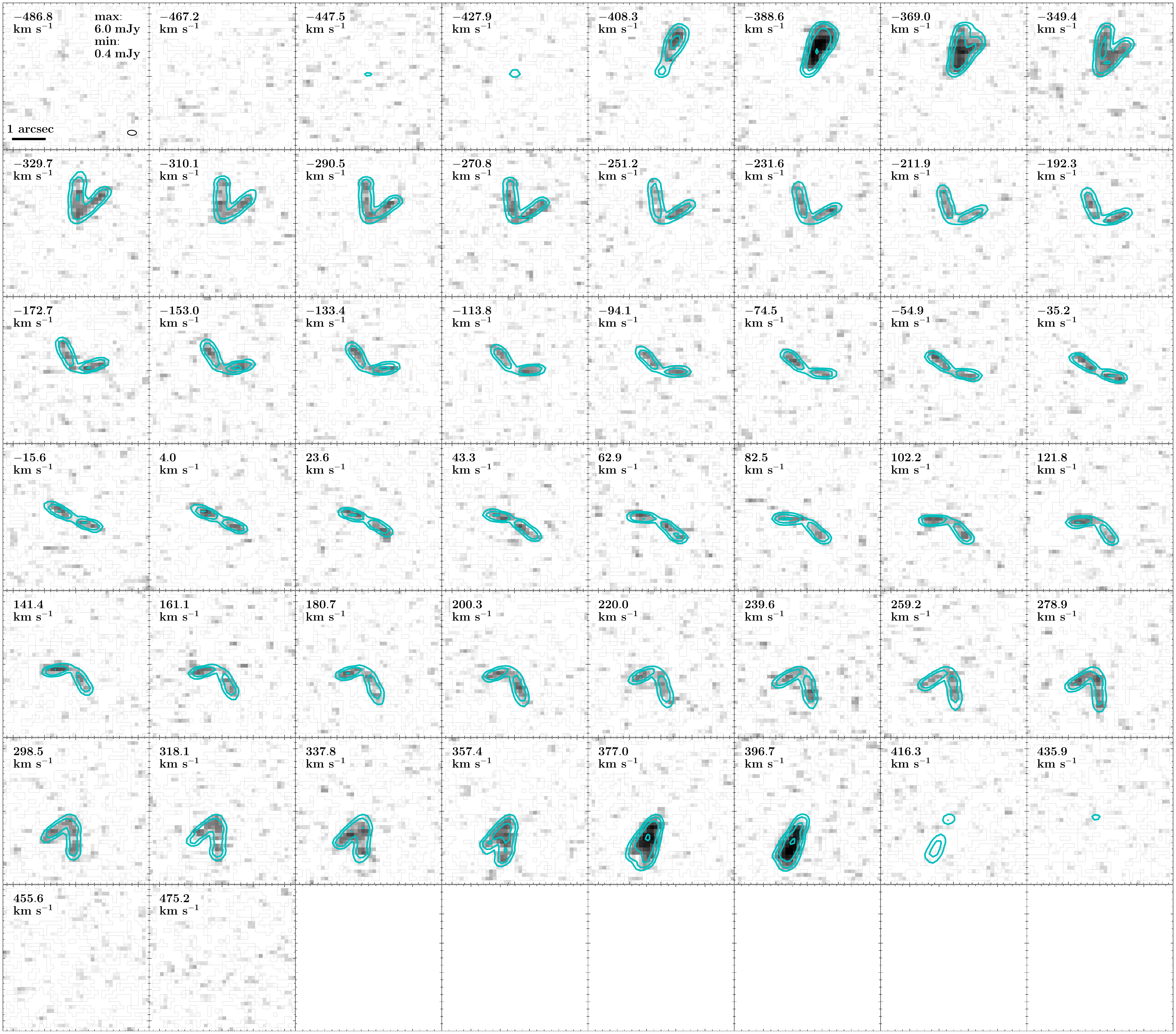} 
\caption{Flux maps for each channel in the data cube, showing the data (black) with the best-fit fiducial model contours overlaid (cyan).
The line-of-sight velocity, with the systemic velocity subtracted, is shown in the top left of each panel.
The minimum and maximum fluxes plotted for both data and model are listed in the top left panel.
The data and model are shown on the same logarithmic scale.
The top left panel also displays a 1\arcsec\ bar and a black ellipse showing the ALMA synthesized beam.
Low levels of model flux extend beyond the data in a few of the highest velocity channels, matching the low-signal upturns seen in Figure \ref{fig_pvds}.
Nevertheless, in general, the model matches the observations quite well.
}
\label{fig_fiducial_channelmaps}
\end{figure*}

\begin{figure*}
\includegraphics[width=\textwidth]{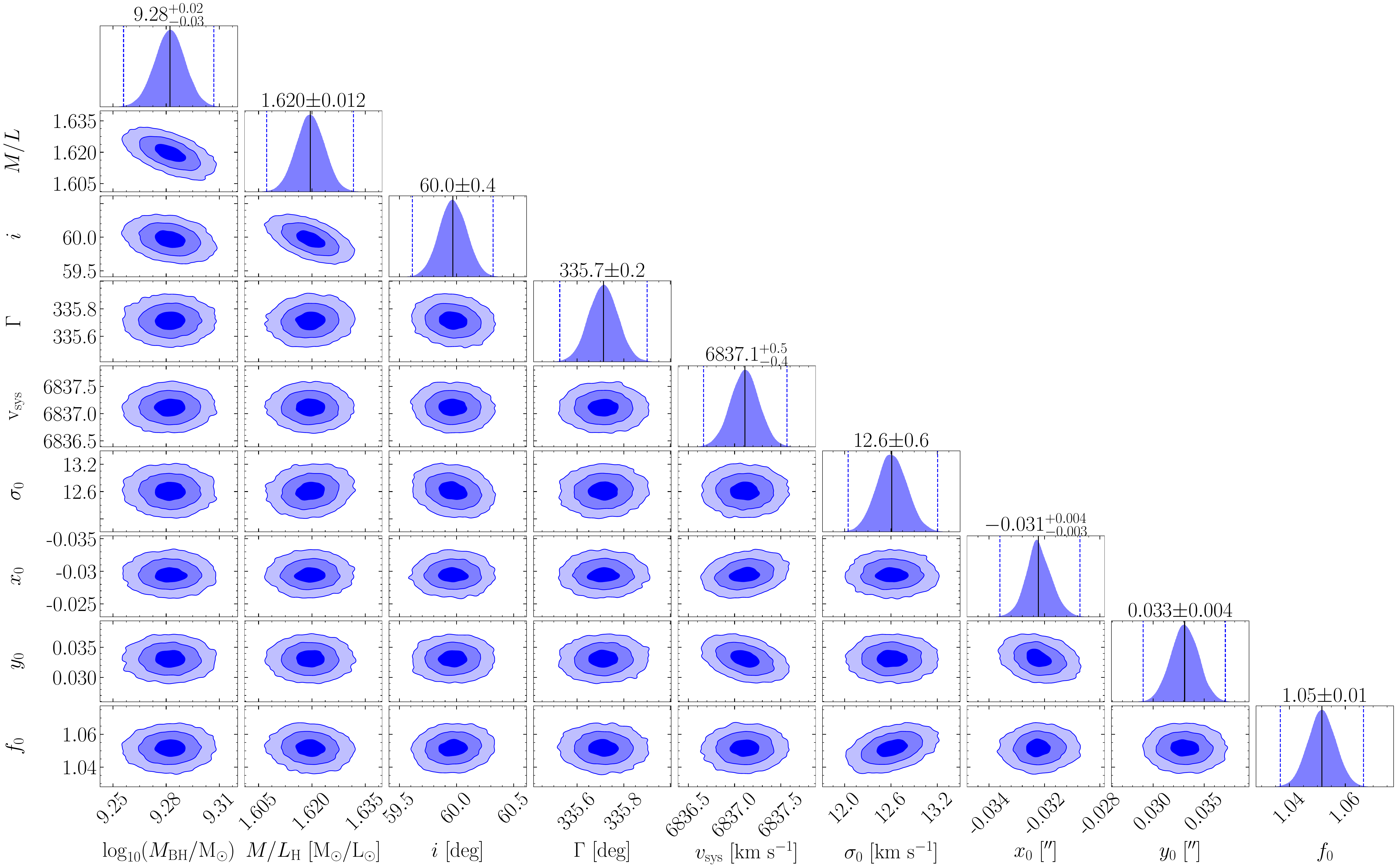}  
\caption{Corner plot of parameters from the fiducial model, showing their one-dimensional (1D; top edge) and 2D posteriors.
Medians and $3\sigma$ confidence intervals are displayed with the 1D posteriors with a black solid line and dashed blue lines, respectively, with values listed above the 1D panels for each parameter.
Contours in the 2D posterior panels correspond to $1\sigma$, $2\sigma$, and $3\sigma$ confidence levels.
Our best-fit BH mass is \mbh\ $= (1.91\pm0.04[\pm0.11]) \times 10^9$ $M_\odot$ ($1\sigma$ [$3\sigma$] statistical uncertainties).
For the best-fit model, we take the median of each parameter's posterior, resulting in $\chi^2_\nu = 1.458$.
}
\label{fig_fiducial_corner_pgc}
\end{figure*}

The moment maps and the PVD constructed from the best-fit model are shown in Figures \ref{fig_fiducial_moments_pgc} and \ref{fig_pvds} on the original ALMA $0\farcs{03}$ pixel$^{-1}$ scale.
We additionally present the residuals between the data and model for the first moment map and the PVD.
Figure \ref{fig_pvds} further includes a PVD constructed from a best-fit model in which \mbh\ was held fixed to 0.
This model's residual indicates a nonzero \mbh\ is required to match the observations.
We show the observed and best-fit model line profiles for every down-sampled pixel in the fitting ellipse in Appendix \ref{appendix}.
As another comparison between the data and best-fit model, we plot the flux for each velocity channel in the data cube overlaid with model flux contours in Figure \ref{fig_fiducial_channelmaps}.
Figures \ref{fig_fiducial_moments_pgc}, \ref{fig_pvds}, \ref{fig_fiducial_channelmaps}, and the line profiles in Appendix \ref{appendix} indicate the best-fit model is a good fit to the data.
As seen in previous work (e.g., \citealt{Cohn2021}), the model modestly underestimates the observed flux near the center.
Previously, it has been shown that the choice of intrinsic flux map can affect the fit quality, but does not significantly affect the inferred \mbh\ \citep{Marconi2006,Walsh2013,Boizelle2021}.
A corner plot displaying the posterior distributions is shown in Figure \ref{fig_fiducial_corner_pgc}.

\begin{deluxetable}{llllc}[t]
\tabletypesize{\small}
\tablecaption{Modeling Results}
\tablewidth{0pt}
\tablehead{
\colhead{Parameter} & 
\colhead{Median} & 
\colhead{$1\sigma$} &
\colhead{$3\sigma$} &
\colhead{Prior range}
\\[-1.5ex]
\colhead{(1)} & 
\colhead{(2)} & 
\colhead{(3)} &
\colhead{(4)} &
\colhead{(5)}
}
\startdata
\mbh\ [$10^9\ M_\odot$] & ${1.91}$ & $\pm0.04$ & $\pm0.11$ & [${0.10}, {10.00}$] \\
\ml\ [$M_\odot/L_\odot$] & ${1.620}$ & $\pm0.004$ & $\pm0.012$ & [${0.500}, {3.000}$] \\
\inc\ [$^\circ$] & ${60.0}$ & $\pm0.1$ & $\pm0.4$ & [${57.9}, {89.9}$] \\
\pa\ [$^\circ$] & ${335.7}$ & $\pm0.1$ & $\pm0.2$ & [${225.0}, {355.0}$] \\
\vsys\ [km s$^{-1}$] & ${6837.1}$ & $_{-0.2}^{+0.1}$ & $_{-0.4}^{+0.5}$ & [${6800.0}, {6900.0}$] \\
\sig\ [km s$^{-1}$] & ${12.6}$ & $\pm0.2$ & $\pm0.6$ & [${0.0}, {20.0}$] \\
\xloc\ [\arcsec] & $-0.031$ & $\pm0.001$ & $_{-0.003}^{+0.004}$ & [$0.079, -0.101$] \\
\yloc\ [\arcsec] & $0.033$ & $\pm0.001$ & $\pm0.004$ & [$-0.073, 0.107$] \\
\fw\ & ${1.05}$ & $\pm0.00$ & $\pm0.01$ & [${0.50}, {1.50}$]
\enddata
\begin{singlespace}
\tablecomments{Results from the best-fit model of PGC 11179.
Column (1) lists the free parameters, column (2) shows the median of the posterior distribution, columns (3) and (4) provide $1\sigma$ and $3\sigma$ statistical uncertainties, and column (5) lists the prior range.
The dust-masked MGE can only be deprojected for inclination angles above $57.9^\circ$, so we use this as the lower bound on the \inc\ prior.
The position angle, \pa, is measured in degrees east of north to the blueshifted side of the disk.
The BH coordinates, \xloc\ and \yloc, are measured in arcseconds relative to the maximum of the continuum emission, which is at RA = $2^{\mathrm{h}} 57^{\mathrm{m}} 33.6707^{\mathrm{s}}$ and Dec = $+5^\circ 58\arcmin 37\farcs{053}$ (J2000).
Positive values of \xloc\ and \yloc\ correspond respectively to shifts to the east and north.
}
\end{singlespace}
\label{tab_fiducial_pgc}
\end{deluxetable}

\subsection{Error Budget \label{error_pgc}}

In addition to the statistical uncertainties from {\tt dynesty}, there are other sources of uncertainty that arise from choices we made when constructing our dynamical model.
We follow \citet{Boizelle2019}, \citet{Cohn2021}, and \citet{Kabasares2022} in performing tests to assess the impact of the modeling assumptions on \mbh\ and \ml.

\textit{Dust extinction.}
In addition to the dust-masked MGE that is used in the fiducial dynamical model, we also examine the dust-corrected MGE using $A_H=0.3$.
Additionally, in place of the original MGE, which results in a poor match to the observed surface brightness distribution, we test a more plausible minimally masked MGE (see Section \ref{mge}).
Using the dust-corrected MGE results in a higher BH mass, with \mbh\ $ = 1.98\times10^9\ M_\odot$ (a 3.4\% increase), a consistent \ml\ of 1.61 $M_\odot/L_\odot$, and $\chi^2_\nu = 1.458$.
A negative shift in \mbh\ is seen when adopting the minimally masked MGE, and there is a 25.8\% decrease to \mbh\ $ = 1.42\times10^9\ M_\odot$.
The \ml\ increases to 1.746 $M_\odot/L_\odot$ and $\chi^2_\nu = 1.453$.

\textit{Radial motion.}
Although the first moment map of PGC 11179 does not show any indication of non-circular motion, we follow \citet{Cohn2021} in employing two toy models to determine whether the data favor any radial motion.
The first model uses a radial velocity term ($v_{\text{rad}}$) that is constant with radius.
We project $v_{\text{rad}}$ into the line-of-sight and add it to $v_{\text{los}}$.
With this model, we find an inflow with $v_{\text{rad}} = -17.8^{+3.4}_{-3.2}$ \kms\ ($3\sigma$ uncertainties).
The best-fit \mbh\ decreases by 2.0\%, \ml\ decreases by 0.6\%, and $\chi^2_\nu=1.442$.

The second model uses a dimensionless factor, $\kappa$, that varies with radius (see, e.g., \citealt{Jeter2019}).
Here, $\kappa$ is multiplied by $v_c$ and projected onto the line-of-sight, and then summed with $v_{\text{los}}$.
In this model, we again find a small inflow with $\kappa=-0.04\pm0.01$ ($3\sigma$ uncertainties), producing a 1.9\% decrease to \mbh, a 0.6\% decrease to \ml, and a $\chi^2_\nu$ of 1.442.
With this $\kappa$, the median radial velocity in the disk is $-17.4$ \kms, consistent with $v_{\text{rad}}$ above.
While our models favor a radial inflow, there are only minor changes to the best-fit parameters.

\textit{Gas mass.}
We run another dynamical model that includes the mass of the molecular gas.
First, we measure the CO surface brightness versus radius within elliptical annuli from the zeroth moment map.
After calculating the projected mass surface densities, we assume a thin disk and integrate \citep{Binney2008} to calculate the contribution of the gas mass to the circular velocity ($v_{c,\text{gas}}$) in the galaxy midplane.
The peak circular velocity is $v_{c,\text{gas}} \sim170$ \kms, which is $\sim$37\% of the maximum circular velocity due to the stars ($\sim$460 \kms).
Nevertheless, the best-fit parameters and uncertainties are nearly unchanged from the fiducial model, and $\chi^2_\nu$ increases to 1.511.

\textit{Turbulent velocity dispersion.}
Our primary dynamical model has a turbulent velocity dispersion that is constant with radius.
We also explore a $\sigma_{\text{turb}}$ that varies exponentially with radius: $\sigma_{\text{turb}} = \sigma_0 + \sigma_1 \exp(-r/r_0)$.
We find $\sigma_0 = 5.0^{+4.0}_{-4.9}$ \kms, with $\sigma_1 = 24.4^{+8.4}_{-5.2}$ \kms\ and $r_0 = 389.1^{+308.6}_{-188.6}$ pc ($3\sigma$ uncertainties), corresponding to $\sim$30.1 ALMA pixels or $\sim$0\farcs{903}.
For this model, \mbh\ falls to $1.81\times10^9\ M_\odot$ (a 5.5\% decrease from the fiducial model), \ml\ increases by 14.3\% to 1.636 $M_\odot/L_\odot$, and $\chi^2_\nu$ improved moderately to $1.444$.

\textit{Oversampling factor.}
Previous ionized (e.g., \citealt{Barth2001}) and molecular (e.g., \citealt{Boizelle2019}) gas-dynamical studies that a low spatial oversampling factor may bias the \mbh\ value, although the impact is not always significant (e.g., $\lesssim 2\%$; \citealt{Cohn2021}).
We use a spatial pixel oversampling of $s = \oversnp$ in the fiducial dynamical model, and also test $s =$1, 2, 3, 4, 8, and 10.
There are very minor changes to \mbh\ between the different oversampling factors.
The model with $s=1$ produces the greatest increase in \mbh\ ($0.51\%$ larger than the fiducial model \mbh), while the $s=3$ model yields the most significant decrease ($0.13\%$ smaller than the fiducial model BH).
The best-fit model using $s=1$ has $\chi^2_\nu=1.461$, and the model with $s=3$ results in $\chi^2_\nu=1.459$, which are both slightly worse than the fiducial $\chi^2_\nu$ of $1.458$.
Both models produce a best-fit \ml\ unchanged from the fiducial model.

\textit{Down-sampling factor.}
In order to account for correlated noise, we down-sample the model and data cubes into groups of $\ds\times\ds$ pixels before calculating the likelihood.
The ALMA beam for PGC 11179 is elongated, with a minor axis FWHM $=0\farcs{16}$ (5.3 pixels) and a major axis FWHM $=0\farcs{29}$ (9.7 pixels).
The major axis of the beam is nearly aligned with the $x$-axis of the PGC 11179 ALMA data cube, with PA $=86.97^\circ$ east of north.
Hence, we explore a down-sampling factor of $10\times5$ spatial pixels.
The model produces $M_\mathrm{BH} = 1.87\times10^9\ M_\odot$ (a $2.4\%$ decrease), \ml$=$1.626 (a 0.4\% increase), and $\chi^2_\nu=1.541$.

\textit{Fitting ellipse.}
The fiducial model from \S \ref{p11179results} uses $r_{\text{fit}} = 1\farcs{6}$, but we further adopt $r_{\text{fit}}$ of $1\farcs{4}$ and $1\farcs{8}$.
For the model with $r_{\text{fit}} = 1\farcs{4}$, we find $M_\mathrm{BH} = 2.01\times10^9\ M_\odot$ (a 4.9\% increase), \ml$=$1.605 (a 0.9\% decrease), and $\chi^2_\nu=1.471$.
The model with $r_{\text{fit}} = 1\farcs{8}$ produces $M_\mathrm{BH} = 1.88\times10^9\ M_\odot$ (a 2.1\% decrease), \ml$=$1.627 (a 0.4\% increase), and $\chi^2_\nu=1.450$.

\textit{Intrinsic flux map.}
Finally, we analyze the impact on the best-fit model parameters of the assumed intrinsic CO flux.
We change the number of iterations of the Lucy-Richardson deconvolution process when constructing the intrinsic flux map used to weight the model line profiles.
Here, we try 5 and 15 iterations, in addition to the 10 iterations employed in our fiducial model.
With 5 and 15 iterations, we find a 0.1\% decrease to \mbh\ and a 0.2\% increase to \mbh, respectively.
The \ml\ decreases by 0.2\% for the model with 5 iterations and is unchanged from the fiducial value for the model with 15 iterations.
The best-fit dynamical model using the former (latter) intrinsic flux map has $\chi^2_\nu=1.449$ ($\chi^2_\nu=1.469$).

\textit{Final error budget}.
We estimate the positive and negative systematic (sys) uncertainties by summing the corresponding changes to \mbh\ and \ml\ described above in quadrature.
This results in systematic uncertainties of 6\% (upper bound) and 27\% (lower bound) on \mbh.
These systematic uncertainties are much larger than the $1\sigma$ statistical (stat) uncertainties of $\sim$2\%.
The BH mass increases the most when we allow the turbulent velocity dispersion to vary spatially, and \mbh\ decreases by a larger amount when we use the minimally masked MGE.
However, the fiducial dust-masked MGE is a better fit to the HST data than the minimally masked MGE is.
The greatest \ml\ increase comes from the model using the exponential turbulent velocity dispersion and the greatest decrease comes from the model with $r_{\text{fit}} = 1\farcs4$.
Ultimately, the BH mass in PGC 11179 is \mbhbothsigpgc.
The \ml\ is $1.620\pm0.004$ [$1\sigma$ stat] $\pm0.012$ [$3\sigma$ stat] $^{+0.211}_{-0.107}$ [sys] $M_\odot/L_\odot$.

\section{Discussion \label{discussion}}

This is the first dynamical BH measurement in PGC 11179, and the second molecular gas-dynamical determination in the local compact galaxy sample from \cite{Yildirim2017}.
Below, we discuss the BH SOI in PGC 11179 (\S \ref{resolution}), the location of the galaxy relative to the BH scaling relations (\S \ref{scaling_relations}), and the implications for our understanding of BH$-$host galaxy co-evolution (\S \ref{bh_gal_growth}).

\subsection{The BH Sphere of Influence \label{resolution}}

We first estimate the BH SOI by calculating the radius where \mbh\ equals the enclosed stellar mass.
For the fiducial model, this occurs at $r_{\text{SOI}}=0\farcs{14}$ (61 pc).
A \textbf{useful} proxy for the BH SOI is $r_{\text{g}} = G$\mbh$/\sigma_\star^2$.
Taking $\sigma_\star = 266$ km s$^{-1}$ \citep{Yildirim2017}, we calculate a larger $r_{\text{g}} = 0\farcs{27}$ (116 pc).
The ALMA beam is $0\farcs{29}\times0\farcs{16}$, with a geometric mean of $0\farcs{22}$.
Comparing the geometric mean of the ALMA beam dimensions to the former (latter) measurement of the BH SOI, $\xi = 2r_\mathrm{SOI}/\theta_\mathrm{FWHM} = 1.3$ (2.4), indicating a marginally resolved BH SOI.
With $\xi$$\sim$2, statistical uncertainties on the BH mass are expected to be small, but it remains vital to account for systematic uncertainties (e.g., \citealt{Davis2014,Cohn2021}).

Our values of $\xi$ are comparable to those for several other ALMA molecular gas-dynamical \mbh\ measurements, which usually range from $\sim$1$-$2 (e.g., \citealt{Barth2016b,Onishi2017,Cohn2021,Davis2017,Davis2018,Nguyen2020,Smith2019,Smith2021,Kabasares2022}), although there are also some measurements with $\xi<1$ (e.g., \citealt{Onishi2015,Thater2019,Nguyen2022}).
Currently, there are only a couple measurements that highly resolve the BH SOI ($\xi>10$; \citealt{Boizelle2019,North2019}).
Relative to our measurements of the CO surface brightness and gas distribution on sub-arcsecond scales, future ALMA observations can be obtained at higher angular resolution to reduce the \mbh\ uncertainties (e.g., \citealt{Boizelle2017,Boizelle2019}).

\subsection{The BH Scaling Relations \label{scaling_relations}}

To locate PGC 11179 relative to the BH scaling relations, we take $\sigma_\star = 266\pm9$ km s$^{-1}$, which is the stellar velocity dispersion within a circular aperture containing half of the galaxy light \citep{Yildirim2017}.
We further sum the components of the dust-masked MGE to calculate the total $H$-band luminosity, then convert to a total $K$-band luminosity assuming an absolute $H$-band ($K$-band) Solar magnitude of 3.37 (3.27) mag \citep{Willmer2018} and $H-K=0.2$ mag from SSP models \citep{Vazdekis2010}.
For PGC 11179, we determine $L_H = 8.72\times10^{10}\ L_\odot$ and $L_K = 9.56\times10^{10}\ L_\odot$.
The galaxy luminosity can be converted to a total stellar mass estimate by multiplying $L_H$ by the best-fit \ml\ from the fiducial model ($1.62\ M_\odot/L_\odot$), resulting in $M_{\star} = 1.41\times10^{11}\ M_\odot$.

\begin{figure*}
\centering
\includegraphics[width=\textwidth]{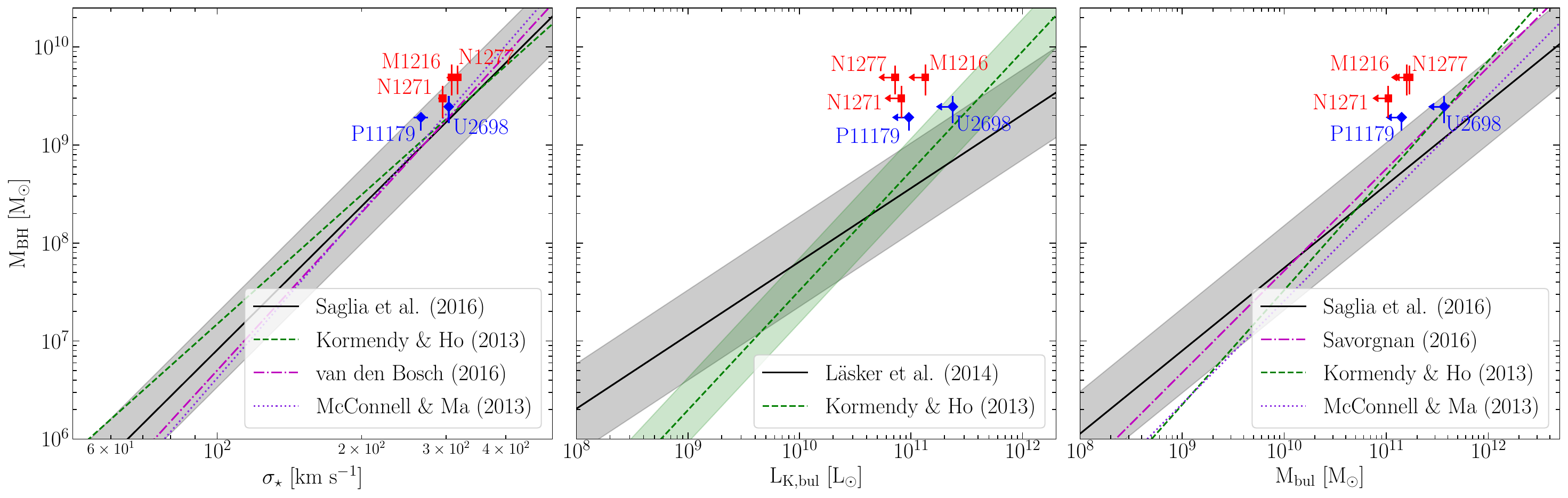}
\caption{Location of the local compact galaxies with dynamical BH masses relative to the \msig\ (left), \mlum\ (middle), and \mmass\ (right) relationships.
The ALMA-based \mbh\ for PGC 11179 measured in this work is shown as a blue diamond, along with the molecular gas-dynamical \mbh\ determination for UGC 2698 \citep{Cohn2021}.
The stellar-dynamical BH masses from integral field spectroscopy assisted by adaptive optics are displayed as red squares \citep{Walsh2015,Walsh2016,Walsh2017}.
The shaded regions indicate the intrinsic scatter of the scaling relations.
All of the galaxies are consistent with \msig.
We conservatively use the total luminosity and mass for each of the compact galaxies when plotting ``bulge" quantities.
As such, these estimates are upper bounds on the bulge measurements.
NGC 1271, NGC 1277, and Mrk 1216, as well as PGC 11179, are positive outliers from the \mlum\ and \mmass\ relations.
However, UGC 2698 is consistent with the \mlum\ and \mmass\ correlations within their intrinsic scatter.}
\label{fig_scaling_relations}
\end{figure*}

In Figure \ref{fig_scaling_relations} we show PGC 11179 on the \msig, \mlum, and \mmass\ relations.
The other local compact galaxies with dynamical \mbh\ measurements -- NGC 1271, NGC 1277, Mrk 1216, and UGC 2698 -- are also plotted, and their host galaxy properties are taken from \cite{Cohn2021}.
Since there has been disagreement on the bulge properties of the local compact galaxy sample \citep{SavorgnanGraham2016,Graham2016b}, we conservatively adopt the total galaxy luminosity and mass when placing the six objects on the BH scaling relations.
For all \mbh\ measurements, we display total uncertainties, calculated as the $1\sigma$ statistical and systematic uncertainties added in quadrature.
We find that the BH in PGC 11179 is over-massive relative to both the \mlum\ and \mmass\ relations but is in agreement with the \msig\ relation.
This result is akin to those for NGC 1271, NGC 1277, and Mrk 1216, although it is a less significant positive outlier, falling a factor of 3.7 and 3.6 above the \mbh\ expected from the \mlum\ and \mmass\ relations, respectively \citep{Kormendy2013, Saglia2016}.

In contrast, \citet{Zhu2021} create new BH scaling relations using only classical bulges and the cores of nearby ellipticals, which are thought to be assembled from $z\sim2$ red nuggets.
Conservatively using the total stellar mass calculated for PGC 11179 as the core mass, the BH mass we measure is only a factor of 1.4 above the \citet{Zhu2021} \mbh$-M_\mathrm{core}$ scaling relation, which is within the scatter of the relation.
The BH mass in UGC 2698 is also within the scatter of this relation, albeit under-massive, while the BH masses in NGC 1271, NGC 1277, and Mrk 1216 remain slight positive outliers (each a factor of $\sim$3 above the relation).

\subsection{BH and Galaxy Growth \label{bh_gal_growth}}

PGC 11179 is a typical object within the local compact galaxy sample from \citet{Yildirim2017}.
It has an effective radius of 1.8 kpc \citep{Yildirim2017} and a stellar mass of $1.41\times10^{11}\ M_\odot$, making it consistent with the $z\sim2$ mass$-$size relation \citep{Wel2014}.
Like the other local compact galaxies, PGC 11179 is a fast rotator with a disk-like shape and a stellar mass surface density that is elevated at the center with a steep decrease at large radii \citep{Yildirim2017}.
PGC 11179 and the other local compact galaxies host uniformly old ($\sim$10 Gyr) stellar populations and super-solar stellar metallicities, consistent with the centers of local giant ellipticals \citep{Trujillo2014,Martin2015,Mateu2017}.
In addition, PGC 11179, along with the other compact galaxies, has a large red (metal-rich) globular cluster population \citep{Beasley2018,Kang2021}.
Thus, the local compact galaxies are distinct from the round, pressure-supported, slow rotating giant ellipticals and BCGs in the nearby Universe.
Instead, it has been posited that the local compact galaxies may be relics of red nugget galaxies that have only evolved passively since $z\sim2$ \citep{Mateu2015,Yildirim2017}.

Most $z\sim2$ red nuggets are thought to undergo dry mergers, forming the cores of massive local elliptical galaxies \citep{Naab2009,Oser2010,Dokkum2010}.
These mergers would increase bulge stellar masses and luminosities without significantly increasing \mbh.
Cosmological hydrodynamical simulations have also reported that some $z\sim2$ red nuggets evolve passively to $z\sim0$ \citep{Wellons2016}.
An alternative evolutionary history for the compact galaxies is that they could have been giant ellipticals that had their outer layers stripped away.
However, the local compact galaxies display regular isophotes and no apparent signs of tidal interactions.
Environmentally, most of the local compact galaxies are found in cluster environments, and PGC 11179 is located in Abell 400 \citep{Beers1992}, although some sample members are isolated field galaxies \citep{Yildirim2017}.

If PGC 11179 and the other local compact galaxies are relics of $z\sim2$ red nuggets, finding that they host over-massive BHs could indicate that BH growth occurs prior to stellar growth in the outskirts of massive galaxies.
Although it is a less significant positive outlier from the BH scaling relations than the local compact galaxies NGC 1271, NGC 1277, and Mrk 1216 \citep{Walsh2015,Walsh2016,Walsh2017}, PGC 11179 is consistent with the picture that $z\sim2$ red nugget relics host over-massive BHs.
The local compact galaxy UGC 2698 challenges this idea due to a recent ALMA-based gas-dynamical \mbh\ measurement.
\cite{Cohn2021} find that UGC 2698 agrees with all of the BH$-$galaxy relations, but UGC 2698 also has properties consistent with being a less pristine relic and perhaps represents an intermediary evolutionary step between the $z\sim2$ red nuggets and the $z\sim0$ massive early-type galaxies \citep{Yildirim2017}.

Our results could also be explained by the existence of greater scatter in the high-mass end of the scaling relations than previously thought, in which case the over-massive BHs may simply be very unusual cases.
However, if that is the case, the reason for such over-massive BHs to appear to be more common in local compact galaxies remains to be explained.
At the moment, it is also difficult to disentangle intrinsic scatter in the scaling relations from scatter due to systematics introduced by the use of different measurement techniques (e.g., stellar-dynamical and molecular gas-dynamical measurements).
To clarify the picture, more precision \mbh\ measurements are required at the upper end of the BH scaling relations, including more BH measurements for the local compact galaxy sample and independent cross-checks between stellar- and molecular gas-dynamical methods.
Such measurements will help to establish whether the sample of local compact galaxies, and plausible $z\sim2$ red nugget relics, host over-massive BHs and whether BH growth predominantly occurs before galaxy growth.

\section{Conclusions\label{conclusions}}

PGC 11179 is a local compact galaxy and possible $z\sim2$ red nugget relic, distinct from common local massive early-type galaxies \citep{Yildirim2017}.
We used ALMA to observe CO($2-1$) emission in the nuclear gas disk of PGC 11179 with $0\farcs{22}$ resolution.
We mapped spatially resolved gas kinematics and fit thin disk dynamical models to the ALMA data, measuring \mbhfullerrpgc\ and stellar \ml\ $=1.620\pm0.004$ [$1\sigma$ stat] $^{+0.211}_{-0.107}$ [sys] $M_\odot/L_\odot$.
The ALMA observations marginally resolve the BH SOI, and the estimated systematic uncertainties on \mbh\ are an order of magnitude larger than the statistical uncertainties.

We find that PGC 11179 it is consistent with the \msig\ relation but lies above \mlum\ and \mmass\ relations.
This matches the behavior of three other local compact galaxies with previous stellar-dynamical \mbh\ measurements \citep{Walsh2015, Walsh2016, Walsh2017}, although PGC 11179 is not as significant of an outlier.
If the local compact galaxies are relics of $z\sim2$ red nuggets and display over-massive BHs, it could indicate that the growth of BHs precedes the growth of stellar spheroids in massive galaxies.
In this scenario, galaxies would typically gain stellar mass in their outskirts via dry mergers to bring them into alignment with the local scaling relations \citep{Martin2015}.

PGC 11179 is the second local compact galaxy with a molecular gas-dynamical \mbh\ measurement from ALMA.
Previously, \cite{Cohn2021} modeled ALMA observations of UGC 2698 and found that the BH was consistent with the \msig, \mlum, and \mmass\ relations.
Unlike PGC 11179, which is more representative of the local compact galaxy sample, UGC 2698 may be a less pristine relic \citep{Yildirim2017} and may have already evolved toward the present-day BH scaling relations.

Alternatively, there may be more scatter in the scaling relations than previously thought.
Systematic differences between stellar-dynamical and molecular gas-dynamical measurement methods can also inflate the scatter in the scaling relations, an effect that is poorly understood due to the very small number of direct comparison studies \citep{Krajnovic2009,Rusli2011,Schulze2011,Rusli2013a,Barth2016a,Davis2017,Smith2019,Kabasares2022}.
More \mbh\ measurements for the sample using stellar-dynamical and molecular gas-dynamical models -- including consistency tests for the same galaxies -- are necessary to clarify the situation.
The high resolution and sensitivity offered by ALMA present the opportunity to make more \mbh\ measurements in the local compact galaxy sample.
Obtaining stellar-dynamical \mbh\ measurements for PGC 11179 and UGC 2698 would also help us understand their behavior relative to the BH scaling relations and provide insights into BH$-$host galaxy co-evolution. 

\section*{Acknowledgements}
J.~H.~C. and J.~L.~W. were supported in part by NSF grant AST-1814799.
This paper makes use of the following ALMA data: ADS/JAO.ALMA\#2016.1.01010.S.
ALMA is a partnership of ESO (representing its member states), NSF (USA), and NINS (Japan), together with NRC (Canada), MOST and ASIAA (Taiwan), and KASI (Republic of Korea), in cooperation with the Republic of Chile.
The Joint ALMA Observatory is operated by ESO, AUI/NRAO, and NAOJ.
The National Radio Astronomy Observatory is a facility of the National Science Foundation operated under cooperative agreement by Associated Universities, Inc.
This paper is based on observations made with the NASA/ESA Hubble Space Telescope, obtained from the data archive at the Space Telescope Science Institute.
The observations are associated with program 13050.
Some of the data presented in this paper were obtained from the Mikulski Archive for Space Telescopes (MAST) at the Space Telescope Science Institute.
The specific observations analyzed can be accessed via \dataset[DOI]{https://doi.org/10.17909/fm1g-y036}.
STScI is operated by the Association of Universities for Research in Astronomy, Inc. under NASA contract NAS 5-26555.
Portions of this research were conducted with the advanced computing resources provided by Texas A\&M High Performance Research Computing.
This work used arXiv.org and NASA's Astrophysics Data System for bibliographic information.
L.~C.~H. was supported by the National Science Foundation of China (11721303, 11991052, 12011540375, 12233001), the National Key R\&D Program of China (2022YFF0503401), and the China Manned Space Project (CMS-CSST-2021-A04, CMS-CSST-2021-A06).

J.~H.~C. would like to thank Silvana Delgado Andrade for help troubleshooting MGE fits and Ryan Hickox for helpful discussions regarding figure design.
The authors thank the anonymous referee for their constructive comments.

\software{CASA (v4.7.2; \citealt{McMullin2007,CASA2022}), \texttt{dynesty} (\citealt{Speagle2020}, \dataset[DOI]{https://doi.org/10.5281/zenodo.7995596}), GALFIT \citep{Peng2010}, kinemetry \citep{Krajnovic2006}, mgefit \citep{Cappellari2002}, Tiny Tim \citep{Krist2004}, AstroDrizzle \citep{Gonzaga2012,Hack2012}, scikit-image \citep{Walt2014}, ASTROPY \citep{Astropy2013,Astropy2018,Astropy2022}, MATPLOTLIB \citep{Hunter2007}, NUMPY \citep{Walt2011,Harris2020}, SCIPY \citep{Virtanen2020}.}

\appendix
\section{Line profiles\label{appendix}}
\begin{figure*}
\includegraphics[width=0.8\textwidth]{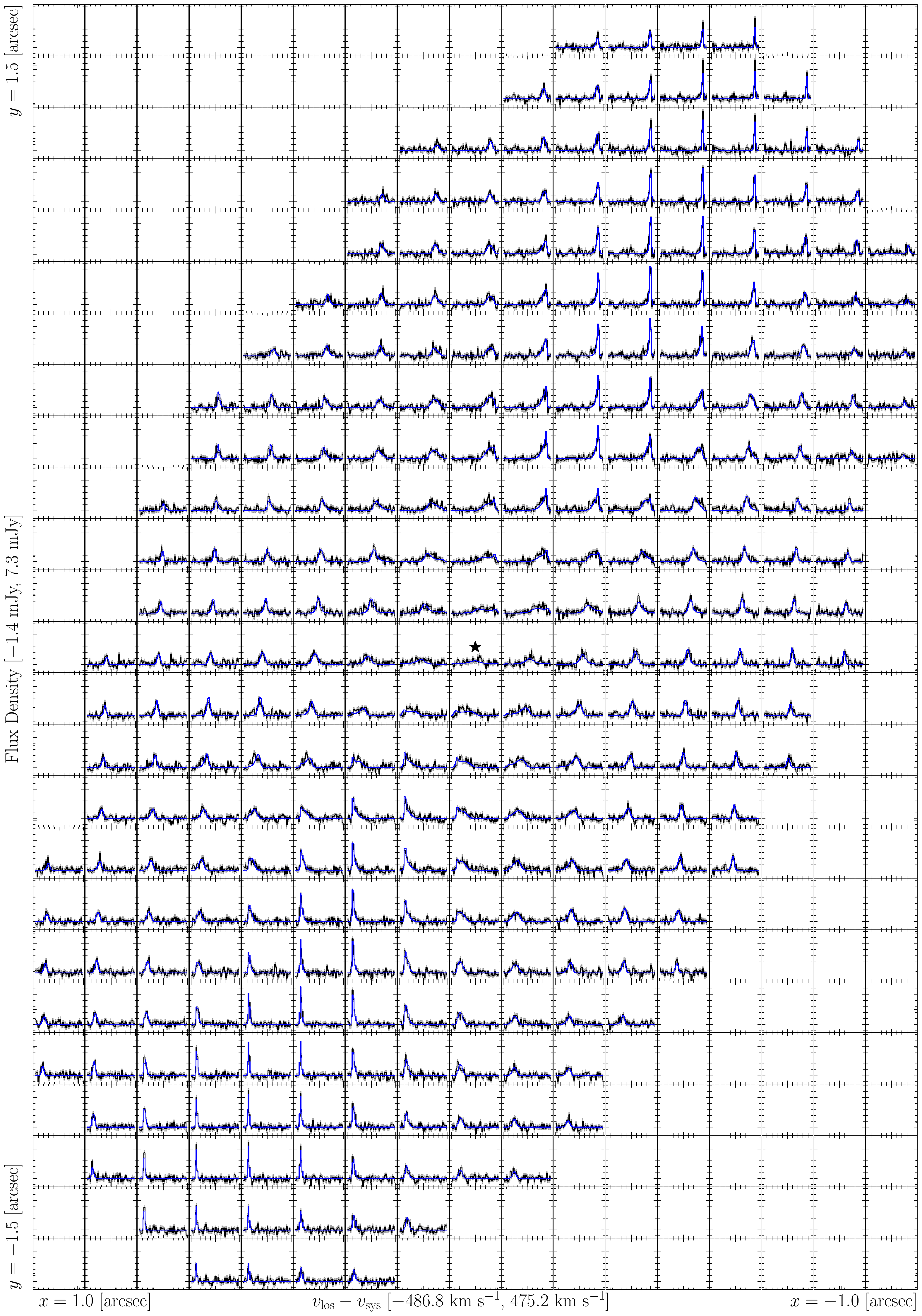} 
\caption{Line profiles built from the down-sampled data cubes (black) and the best-fit model cubes (blue) at each pixel location in the fitting ellipse region.
The noise in each channel is displayed as a shaded gray region.
The velocity and flux ranges of each line profile panel is labeled on the x and y-axis, respectively.
The physical extent of the fitting region is also labeled on each axis.
The black star labels the best-fit BH location.
The model matches the observations quite well throughout the disk.
}
\label{fig_fiducial_lps}
\end{figure*}

As described in \S\ref{model}, the model cube is fit directly to the data cube by comparing the line profiles in each down-sampled pixel in the elliptical fitting region.
To illustrate this process, we plot every observed and best-fit fiducial model line profile in the fitting ellipse in Figure \ref{fig_fiducial_lps}.
This figure indicates that the model is a good fit to the data throughout the disk of PGC 11179.

\begin{thebibliography}{40}

\bibitem[Astropy Collaboration et al.(2013)]{Astropy2013} Astropy Collaboration, Robitaille, T.~P., Tollerud, E.~J., et al.\ 2013, \aap, 558, A33

\bibitem[Astropy Collaboration et al.(2018)]{Astropy2018} Astropy Collaboration, Price-Whelan, A.~M., Sip{\H{o}}cz, B.~M., et al.\ 2018, \aj, 156, 123

\bibitem[Astropy Collaboration et al.(2022)]{Astropy2022} Astropy Collaboration, Price-Whelan, A.~M., Lim, P.~L., et al.\ 2022, \apj, 935, 167. doi:10.3847/1538-4357/ac7c74

\bibitem[Barth et al.(2001)]{Barth2001} Barth, A.~J., Sarzi, M., Rix, H.-W., et al.\ 2001, \apj, 555, 685

\bibitem[{Barth {et~al.}(2016)Barth, Boizelle, Darling, Baker, Buote, Ho, \& Walsh}]{Barth2016a}
Barth, A.~J., Boizelle, B.~D., Darling, J., {et~al.} 2016, \apjl, 822, L2

\bibitem[{Barth {et~al.}(2016b)Barth, Darling, Baker, Boizelle, Buote, Ho, \& Walsh}]{Barth2016b}
Barth, A.~J., Darling, J., Baker, A.~J., {et~al.} 2016, \apj, 823, 51

\bibitem[Beasley et al.(2018)]{Beasley2018} Beasley, M.~A., Trujillo, I., Leaman, R., et al.\ 2018, \nat, 555, 483

\bibitem[Beers et al.(1992)]{Beers1992} Beers, T.~C., Gebhardt, K., Huchra, J.~P., et al.\ 1992, \apj, 400, 410. doi:10.1086/172006

\bibitem[Binney \& Tremaine(2008)]{Binney2008} Binney, J. \& Tremaine, S.\ 2008, Galactic Dynamics: Second Edition, by James Binney and Scott Tremaine. ISBN 978-0-691-13026-2 (HB). Published by Princeton University Press, Princeton, NJ USA, 2008.

\bibitem[Boizelle et al.(2017)]{Boizelle2017} Boizelle, B.~D., Barth, A.~J., Darling, J., et al.\ 2017, \apj, 845, 170

\bibitem[Boizelle et al.(2019)]{Boizelle2019} Boizelle, B.~D., Barth, A.~J., Walsh, J.~L., et al.\ 2019, \apj, 881, 10

\bibitem[Boizelle et al.(2021)]{Boizelle2021} Boizelle, B.~D., Walsh, J.~L., Barth, A.~J., et al.\ 2021, \apj, 908, 19. doi:10.3847/1538-4357/abd24d

\bibitem[Briggs(1995)]{Briggs1995} Briggs, D.~S.\ 1995, American Astronomical Society Meeting Abstracts

\bibitem[Buote \& Barth(2018)]{Buote2018} Buote, D.~A. \& Barth, A.~J.\ 2018, \apj, 854, 143. doi:10.3847/1538-4357/aaa971

\bibitem[Buote \& Barth(2019)]{Buote2019} Buote, D.~A. \& Barth, A.~J.\ 2019, \apj, 877, 91. doi:10.3847/1538-4357/ab1008

\bibitem[Cappellari(2002)]{Cappellari2002} Cappellari, M.\ 2002, \mnras, 333, 400

\bibitem[Carilli \& Walter(2013)]{Carilli2013} Carilli, C.~L. \& Walter, F.\ 2013, \araa, 51, 105

\bibitem[CASA Team et al.(2022)]{CASA2022} CASA Team, Bean, B., Bhatnagar, S., et al.\ 2022, \pasp, 134, 114501. doi:10.1088/1538-3873/ac9642

\bibitem[Cohn et al.(2021)]{Cohn2021} Cohn, J.~H., Walsh, J.~L., Boizelle, B.~D., et al.\ 2021, \apj, 919, 77. doi:10.3847/1538-4357/ac0f78

\bibitem[Davis(2014)]{Davis2014} Davis, T.~A.\ 2014, \mnras, 443, 911

\bibitem[Davis et al.(2017)]{Davis2017} Davis, T.~A., Bureau, M., Onishi, K., et al.\ 2017, \mnras, 468, 4675. doi:10.1093/mnras/stw3217

\bibitem[Davis et al.(2018)]{Davis2018} Davis, T.~A., Bureau, M., Onishi, K., et al.\ 2018, \mnras, 473, 3818. doi:10.1093/mnras/stx2600

\bibitem[Emsellem et al.(1994)]{Emsellem1994} Emsellem, E., Monnet, G., \& Bacon, R.\ 1994, \aap, 285, 723

\bibitem[Emsellem(2013)]{Emsellem2013} Emsellem, E.\ 2013, \mnras, 433, 1862. doi:10.1093/mnras/stt840

\bibitem[Faber et al.(1997)]{Faber1997} Faber, S.~M., Tremaine, S., Ajhar, E.~A., et al.\ 1997, \aj, 114, 1771. doi:10.1086/118606

\bibitem[Ferrarese \& Merritt(2000)]{Ferrarese2000} Ferrarese, L. \& Merritt, D.\ 2000, \apjl, 539, L9. doi:10.1086/312838

\bibitem[Ferr{\'e}-Mateu et al.(2015)]{Mateu2015} Ferr{\'e}-Mateu, A., Mezcua, M., Trujillo, I., et al.\ 2015, \apj, 808, 79

\bibitem[Ferr{\'e}-Mateu et al.(2017)]{Mateu2017} Ferr{\'e}-Mateu, A., Trujillo, I., Mart{\'\i}n-Navarro, I., et al.\ 2017, \mnras, 467, 1929

\bibitem[Fomalont et al.(2014)]{Fomalont2014} Fomalont, E., van Kempen, T., Kneissl, R., et al.\ 2014, The Messenger, 155, 19

\bibitem[Gebhardt et al.(2000)]{Gebhardt2000} Gebhardt, K., Bender, R., Bower, G., et al.\ 2000, \apjl, 539, L13. doi:10.1086/312840

\bibitem[Gonzaga et al.(2012)]{Gonzaga2012} Gonzaga, S. \& et al.\ 2012, The DrizzlePac Handbook, HST Data Handbook

\bibitem[Graham et al.(2016a)]{Graham2016a} Graham, A.~W., Durr{\'e}, M., Savorgnan, G.~A.~D., et al.\ 2016, \apj, 819, 43. doi:10.3847/0004-637X/819/1/43

\bibitem[Graham et al.(2016b)]{Graham2016b} Graham, A.~W., Ciambur, B.~C., \& Savorgnan, G.~A.~D.\ 2016, \apj, 831, 132. doi:10.3847/0004-637X/831/2/132

\bibitem[G{\"u}ltekin et al.(2009)]{Gultekin2009} G{\"u}ltekin, K., Richstone, D.~O., Gebhardt, K., et al.\ 2009, \apj, 698, 198. doi:10.1088/0004-637X/698/1/198

\bibitem[Hack et al.(2012)]{Hack2012} Hack, W.~J., Dencheva, N., Fruchter, A.~S., et al.\ 2012, \aas

\bibitem[Harris et al.(2020)]{Harris2020} Harris, C.~R., Jarrod Millman, K., van der Walt, S.~J., et al.\ 2020, arXiv:2006.10256

\bibitem[Hilz et al.(2013)]{Hilz2013} Hilz, M., Naab, T., \& Ostriker, J.~P.\ 2013, \mnras, 429, 2924. doi:10.1093/mnras/sts501

\bibitem[Hunter(2007)]{Hunter2007} Hunter, J.~D.\ 2007, Computing in Science and Engineering, 9, 90

\bibitem[Jeter et al.(2019)]{Jeter2019} Jeter, B., Broderick, A.~E., \& McNamara, B.~R.\ 2019, \apj, 882, 82. doi:10.3847/1538-4357/ab3221

\bibitem[Kabasares et al.(2022)]{Kabasares2022} Kabasares, K.~M., Barth, A.~J., Buote, D.~A., et al.\ 2022, \apj, 934, 162. doi:10.3847/1538-4357/ac7a38

\bibitem[Kang \& Lee(2021)]{Kang2021} Kang, J. \& Lee, M.~G.\ 2021, \apj, 914, 20. doi:10.3847/1538-4357/abf433

\bibitem[Kormendy \& Ho(2013)]{Kormendy2013} Kormendy, J., \& Ho, L.~C.\ 2013, \araa, 51, 511

\bibitem[Kormendy \& Richstone(1995)]{Kormendy1995} Kormendy, J. \& Richstone, D.\ 1995, \araa, 33, 581. doi:10.1146/annurev.aa.33.090195.003053

\bibitem[Krajnovi{\'c} et al.(2006)]{Krajnovic2006} Krajnovi{\'c}, D., Cappellari, M., de Zeeuw, P.~T., et al.\ 2006, \mnras, 366, 787

\bibitem[Krajnovi{\'c} et al.(2009)]{Krajnovic2009} Krajnovi{\'c}, D., McDermid, R.~M., Cappellari, M., et al.\ 2009, \mnras, 399, 1839. doi:10.1111/j.1365-2966.2009.15415.x

\bibitem[Krajnovi{\'c} et al.(2018)]{Krajnovic2018} Krajnovi{\'c}, D., Cappellari, M., McDermid, R.~M., et al.\ 2018, \mnras, 477, 3030. doi:10.1093/mnras/sty778

\bibitem[Krist \& Hook(2004)]{Krist2004} Krist, J., \& Hook, R. 2004, The Tiny Tim User’s Guide,
\url{http://www.stsci.edu/hst/observatory/focus/TinyTim}, Baltimore:
STScI

\bibitem[Kroupa(2001)]{Kroupa2001} Kroupa, P.\ 2001, \mnras, 322, 231. doi:10.1046/j.1365-8711.2001.04022.x

\bibitem[La Barbera et al.(2019)]{LaBarbera2019} La Barbera, F., Vazdekis, A., Ferreras, I., et al.\ 2019, \mnras, 489, 4090. doi:10.1093/mnras/stz2192

\bibitem[L{\"a}sker et al.(2014)]{Lasker2014} L{\"a}sker, R., Ferrarese, L., van de Ven, G., et al.\ 2014, \apj, 780, 70 

\bibitem[Lavezzi et al.(1999)]{Lavezzi1999} Lavezzi, T.~E., Dickey, J.~M., Casoli, F., et al.\ 1999, \aj, 117, 1995

\bibitem[Lucy(1974)]{Lucy1974} Lucy, L.~B.\ 1974, \aj, 79, 745

{Macchetto}, F., {Marconi}, A., {Axon}, D.~J., {Capetti}, A., {Sparks}, W., \& {Crane}, P. 1997, \apj, 489, 579

\bibitem[Marconi \& Hunt(2003)]{Marconi2003} Marconi, A. \& Hunt, L.~K.\ 2003, \apjl, 589, L21. doi:10.1086/375804

\bibitem[Marconi et al.(2006)]{Marconi2006} Marconi, A., Pastorini, G., Pacini, F., et al.\ 2006, \aap, 448, 921. doi:10.1051/0004-6361:20053853

\bibitem[Mart{\'\i}n-Navarro et al.(2015a)]{Martin2015a} Mart{\'\i}n-Navarro, I., La Barbera, F., Vazdekis, A., et al.\ 2015, \mnras, 447, 1033. doi:10.1093/mnras/stu2480

\bibitem[Mart{\'\i}n-Navarro et al.(2015b)]{Martin2015} Mart{\'\i}n-Navarro, I., La Barbera, F., Vazdekis, A., et al.\ 2015, \mnras, 451, 1081, doi:10.1093/mnras/stv1022

\bibitem[McConnell \& Ma(2013)]{McConnellMa2013} McConnell, N.~J. \& Ma, C.-P.\ 2013, \apj, 764, 184 

\bibitem[McMullin et al.(2007)]{McMullin2007} McMullin, J.~P., Waters, B., Schiebel, D., et al.\ 2007, Astronomical Data Analysis Software and Systems XVI, 376, 127

\bibitem[Mould et al.(2000)]{Mould2000} Mould, J.~R., Huchra, J.~P., Freedman, W.~L., et al.\ 2000, \apj, 529, 786

\bibitem[Naab et al.(2009)]{Naab2009} Naab, T., Johansson, P.~H., \& Ostriker, J.~P.\ 2009, \apjl, 699, L178. doi:10.1088/0004-637X/699/2/L178

\bibitem[Nagai et al.(2019)]{Nagai2019} Nagai, H., Onishi, K., Kawakatu, N., et al.\ 2019, \apj, 883, 193. doi:10.3847/1538-4357/ab3e6e

\bibitem[Neumayer et al.(2007)]{Neumayer2007} Neumayer, N., Cappellari, M., Reunanen, J., et al.\ 2007, \apj, 671, 1329. doi:10.1086/523039

\bibitem[Nguyen et al.(2020)]{Nguyen2020} Nguyen, D.~D., den Brok, M., Seth, A.~C., et al.\ 2020, \apj, 892, 68

\bibitem[Nguyen et al.(2022)]{Nguyen2022} Nguyen, D.~D., Bureau, M., Thater, S., et al.\ 2022, \mnras, 509, 2920. doi:10.1093/mnras/stab3016

\bibitem[North et al.(2019)]{North2019} North, E.~V., Davis, T.~A., Bureau, M., et al.\ 2019, \mnras, 490, 319. doi:10.1093/mnras/stz2598

\bibitem[Onishi et al.(2015)]{Onishi2015} Onishi, K., Iguchi, S., Sheth, K., et al.\ 2015, \apj, 806, 39. doi:10.1088/0004-637X/806/1/39

\bibitem[Onishi et al.(2017)]{Onishi2017} Onishi, K., Iguchi, S., Davis, T.~A., et al.\ 2017, \mnras, 468, 4663. doi:10.1093/mnras/stx631

\bibitem[Oser et al.(2010)]{Oser2010} Oser, L., Ostriker, J.~P., Naab, T., et al.\ 2010, \apj, 725, 2312. doi:10.1088/0004-637X/725/2/2312

\bibitem[Peng et al.(2010)]{Peng2010} Peng, C.~Y., Ho, L.~C., Impey, C.~D., et al.\ 2010, \aj, 139, 2097

\bibitem[Richardson(1972)]{Richardson1972} Richardson, W.~H.\ 1972, Journal of the Optical Society of America (1917-1983), 62, 55

\bibitem[Rieke \& Lebofsky(1985)]{Rieke1985} Rieke, G.~H. \& Lebofsky, M.~J.\ 1985, \apj, 288, 618. doi:10.1086/162827

\bibitem[Ruffa et al.(2019)]{Ruffa2019} Ruffa, I., Prandoni, I., Laing, R.~A., et al.\ 2019, \mnras, 484, 4239. doi:10.1093/mnras/stz255

\bibitem[Ruffa et al.(2023)]{Ruffa2023} Ruffa, I., Davis, T.~A., Cappellari, M., et al.\ 2023, \mnras, 522, 6170. doi:10.1093/mnras/stad1119

\bibitem[Rusli et al.(2011)]{Rusli2011} Rusli, S.~P., Thomas, J., Erwin, P., et al.\ 2011, \mnras, 410, 1223. doi:10.1111/j.1365-2966.2010.17610.x

\bibitem[Rusli et al.(2013)]{Rusli2013a} Rusli, S.~P., Thomas, J., Saglia, R.~P., et al.\ 2013, \aj, 146, 45

\bibitem[Saglia et al.(2016)]{Saglia2016} Saglia, R.~P., Opitsch, M., Erwin, P., et al.\ 2016, \apj, 818, 47

\bibitem[Sandstrom et al.(2013)]{Sandstrom2013} Sandstrom, K.~M., Leroy, A.~K., Walter, F., et al.\ 2013, \apj, 777, 5

\bibitem[Savorgnan et al.(2016)]{Savorgnan2016} Savorgnan, G.~A.~D., Graham, A.~W., Marconi, A., et al.\ 2016, \apj, 817, 21. doi:10.3847/0004-637X/817/1/21 

\bibitem[Savorgnan \& Graham(2016)]{SavorgnanGraham2016} Savorgnan, G.~A.~D. \& Graham, A.~W.\ 2016, \mnras, 457, 320. doi:10.1093/mnras/stv2713

\bibitem[Scharw{\"a}chter et al.(2013)]{Scharwachter2013} Scharw{\"a}chter, J., McGregor, P.~J., Dopita, M.~A., et al.\ 2013, \mnras, 429, 2315. doi:10.1093/mnras/sts502

\bibitem[Scharw{\"a}chter et al.(2016)]{Scharwachter2016} Scharw{\"a}chter, J., Combes, F., Salom{\'e}, P., et al.\ 2016, \mnras, 457, 4272. doi:10.1093/mnras/stw183

\bibitem[Schlafly \& Finkbeiner(2011)]{Schlafly2011} Schlafly, E.~F. \& Finkbeiner, D.~P.\ 2011, \apj, 737, 103

\bibitem[Schulze \& Gebhardt(2011)]{Schulze2011} Schulze, A. \& Gebhardt, K.\ 2011, \apj, 729, 21. doi:10.1088/0004-637X/729/1/21

\bibitem[Seth et al.(2010)]{Seth2010} Seth, A.~C., Cappellari, M., Neumayer, N., et al.\ 2010, \apj, 714, 713. doi:10.1088/0004-637X/714/1/713

\bibitem[Smith et al.(2019)]{Smith2019} Smith, M.~D., Bureau, M., Davis, T.~A., et al.\ 2019, \mnras, 485, 4359. doi:10.1093/mnras/stz625

\bibitem[Smith et al.(2021)]{Smith2021} Smith, M.~D., Bureau, M., Davis, T.~A., et al.\ 2021, \mnras, 503, 5984. doi:10.1093/mnras/stab791

\bibitem[Speagle(2020)]{Speagle2020} Speagle, J.~S.\ 2020, \mnras, 493, 3132. doi:10.1093/mnras/staa278

\bibitem[Thater(2019)]{Thater2019} Thater, S.\ 2019, ALMA2019: Science Results and Cross-Facility Synergies, 129. doi:10.5281/zenodo.3585459

\bibitem[Trujillo et al.(2014)]{Trujillo2014} Trujillo, I., Ferr{\'e}-Mateu, A., Balcells, M., et al.\ 2014, \apjl, 780, L20

\bibitem[van den Bosch et al.(2012)]{Bosch2012} van den Bosch, R.~C.~E., Gebhardt, K., G{\"u}ltekin, K., et al.\ 2012, \nat, 491, 729

\bibitem[van den Bosch et al.(2015)]{Bosch2015} van den Bosch, R.~C.~E., Gebhardt, K., G{\"u}ltekin, K., et al.\ 2015, \apjs, 218, 10

\bibitem[van den Bosch(2016)]{Bosch2016} van den Bosch, R.~C.~E.\ 2016, \apj, 831, 134 

\bibitem[van der Walt et al.(2011)]{Walt2011} van der Walt, S., Colbert, S.~C., \& Varoquaux, G.\ 2011, Computing in Science and Engineering, 13, 22

\bibitem[van der Walt et al.(2014)]{Walt2014}
van der Walt, S., Sch\"onberger, J., Nunez-Iglesias, J, et al.\ 2014, PeerJ, 2, e453

\bibitem[van der Wel et al.(2014)]{Wel2014} van der Wel, A., Franx, M., van Dokkum, P.~G., et al.\ 2014, \apj, 788, 28. doi:10.1088/0004-637X/788/1/28

\bibitem[van Dokkum et al.(2010)]{Dokkum2010} van Dokkum, P.~G., Whitaker, K.~E., Brammer, G., et al.\ 2010, \apj, 709, 1018

\bibitem[Vazdekis et al.(2010)]{Vazdekis2010} Vazdekis, A., S{\'a}nchez-Bl{\'a}zquez, P., Falc{\'o}n-Barroso, J., et al.\ 2010, \mnras, 404, 1639

\bibitem[Viaene et al.(2017)]{Viaene2017} Viaene, S., Sarzi, M., Baes, M., et al.\ 2017, \mnras, 472, 1286. doi:10.1093/mnras/stx1781

\bibitem[Virtanen et al.(2020)]{Virtanen2020} Virtanen, P., Gommers, R., Oliphant, T.~E., et al.\ 2020, Nature Methods, 17, 261

\bibitem[Walsh et al.(2010)]{Walsh2010} Walsh, J.~L., Barth, A.~J., \& Sarzi, M.\ 2010, \apj, 721, 762

\bibitem[Walsh et al.(2013)]{Walsh2013} {Walsh}, J.~L., {Barth}, A.~J., {Ho}, L.~C, \& {Sarzi}, M.\ 2013, \apj, 770, 86

\bibitem[Walsh et al.(2015)]{Walsh2015} Walsh, J.~L., van den Bosch, R.~C.~E., Gebhardt, K., et al.\ 2015, \apj, 808, 183. doi:10.1088/0004-637X/808/2/18

\bibitem[Walsh et al.(2016)]{Walsh2016} Walsh, J.~L., van den Bosch, R.~C.~E., Gebhardt, K., et al.\ 2016, \apj, 817, 2

\bibitem[Walsh et al.(2017)]{Walsh2017} Walsh, J.~L., van den Bosch, R.~C.~E., Gebhardt, K., et al.\ 2017, \apj, 835, 208

\bibitem[Wellons et al.(2016)]{Wellons2016} Wellons, S., Torrey, P., Ma, C.-P., et al.\ 2016, \mnras, 456, 1030

\bibitem[Willmer(2018)]{Willmer2018} Willmer, C.~N.~A.\ 2018, \apjs, 236, 47

\bibitem[Wilman et al.(2005)]{Wilman2005} Wilman, R.~J., Edge, A.~C., \& Johnstone, R.~M.\ 2005, \mnras, 359, 755. doi:10.1111/j.1365-2966.2005.08956.x

\bibitem[Y{\i}ld{\i}r{\i}m et al.(2015)]{Yildirim2015} Y{\i}ld{\i}r{\i}m, A., van den Bosch, R.~C.~E., van de Ven, G., et al.\ 2015, \mnras, 452, 1792. doi:10.1093/mnras/stv1381

\bibitem[{Y\i ld\i r\i m {et~al.}(2017)Y\i ld\i r\i m, Bosch, Ven, Mart\'{i}n-Navarro, Walsh, Husemann, G\"{u}ltekin, \& Gebhardt}]{Yildirim2017}
Y\i ld\i r\i m, A., van den Bosch, R.~E.~C., van den Ven, G., {et~al.} 2017, \mnras, 468, L4216

\bibitem[Zhu et al.(2021)]{Zhu2021} Zhu, P., Ho, L.~C., \& Gao, H.\ 2021, \apj, 907, 6. doi:10.3847/1538-4357/abcaa1

\end{thebibliography}
\end{document}